\documentclass[prb,twocolumn,showpacs,superscriptaddress,preprintnumbers,amssymb]{revtex4}
\usepackage{graphicx}
\usepackage{dcolumn}
\usepackage{bm}

\newcommand{\beq}{\begin{equation}}
\newcommand{\eeq}{\end{equation}}
\newcommand{\beqn}{\begin{eqnarray}}
\newcommand{\eeqn}{\end{eqnarray}}

\newcommand{\ra}{\rightarrow}

\begin{document}

\title{Classification and Description of Bosonic Symmetry Protected Topological
Phases with semiclassical Nonlinear Sigma models}

\author{Zhen Bi}

\author{Alex Rasmussen}

\author{Kevin Slagle}

\author{Cenke Xu}

\affiliation{Department of physics, University of California,
Santa Barbara, CA 93106, USA}

\begin{abstract}

In this paper we systematically classify and describe bosonic
symmetry protected topological (SPT) phases in all physical
spatial dimensions using semiclassical nonlinear Sigma model
(NLSM) field theories. All the SPT phases on a $d-$dimensional
lattice discussed in this paper can be described by the {\it same}
NLSM, which is an O($d+2$) NLSM in $(d+1)-$dimensional space-time,
with a topological $\Theta-$term. The field in the NLSM is a
semiclassical Landau order parameter with a unit length
constraint. The classification of SPT phases discussed in this
paper based on their NLSMs is consistent with the more
mathematical classification based on group cohomology given in
Ref.~\onlinecite{wenspt,wenspt2}. Besides the classification, the
formalism used in this paper also allows us to explicitly discuss
the physics at the boundary of the SPT phases, and it reveals the
relation between SPT phases with different symmetries. For
example, it gives many of these SPT states a natural ``decorated
defect" construction.

\end{abstract}

\date{\today}

\maketitle

\section{Introduction}

Symmetry protected topological (SPT) phase is a new type of
quantum disordered phase. It is intrinsically different from a
trivial direct product state, when and only when the system has
certain symmetry $G$. In terms of its phenomena, a SPT phase on a
$d-$dimensional lattice should satisfy at least the following
three criteria:

({\it i}). On a $d-$dimensional lattice without boundary, this
phase is fully gapped, and nondegenerate;

({\it ii}). On a $d-$dimensional lattice with a
$(d-1)-$dimensional boundary, if the Hamiltonian of the entire
system (including both bulk and boundary Hamiltonian) preserves
certain symmetry $G$, this phase is either gapless, or gapped but
degenerate.


({\it iii}). The boundary state of this $d-$dimensional system
cannot be realized as a $(d-1)$-dimensional lattice system with
the same symmetry $G$.

Both the $2d$ quantum spin Hall
insulator~\cite{kane2005a,kane2005b,bernevig2006} and $3d$
Topological insulator~\cite{fukane,moorebalents2007,roy2007} are
perfect examples of SPT phases protected by time-reversal symmetry
and charge U(1) symmetry. In this paper we will focus on bosonic
SPT phases. Unlike fermion systems, bosonic SPT phases are always
strongly interacting phases of boson systems.

Notice that the second criterion ({\it ii}) implies the following
two possibilities: On a lattice with a boundary, the system is
either gapless, or gapped but degenerate. For example, without
interaction, the boundaries of $2d$ QSH insulator and $3d$ TBI are
both gapless; but with interaction, the edge states of 2$d$ QSH
insulator, and 3$d$ TBI can both be gapped out through spontaneous
time-reversal symmetry breaking at the boundary, and this
spontaneous time-reversal symmetry breaking can occur through a
boundary transition, without destroying the bulk
state~\cite{xuedge,wuedge,xu3dedge}. When $d \geq 3$, the
degeneracy of the boundary can correspond to either spontaneous
breaking of $G$, or correspond to certain topological degeneracy
at the boundary. Which case occurs in the system will depend on
the detailed Hamiltonian at the boundary of the system. For
example, with strong interaction, the boundary of a 3d TBI can be
driven into a nontrivial topological
phase~\cite{wangsenthilTI,qinayakTI,fisherTI,chenTI}.

The concept of SPT phase was pioneered by Wen and his colleagues.
A mathematical paradigm was developed in
Ref.~\onlinecite{wenspt,wenspt2} that systematically classified
SPT phases based on the group cohomology of their symmetry $G$.
But this approach was unable to reveal all the physical properties
of the SPT phases. In the last few years, SPT phase has rapidly
developed into a very active and exciting
field~\cite{wenspt,wenspt2,levingu,levinsenthil,levinstern,liuwen,luashvin,senthilashvin,xu3dspt,xu2dspt,xusenthil,wangsenthil,wangpottersenthil,chenluashvin,maxfisher,yewen1,yewen2,chenggu},
and besides the general mathematical classification, other
approaches of understanding SPT phases were also taken. In 2d, it
was demonstrated that the SPT phases can be thoroughly classified
by the Chern-Simons field theory~\cite{luashvin}, although it is
unclear how to generalize this approach to 3d. Nonlinear Sigma
model field theories were also used to describe some SPT phases in
3d and 2d~\cite{xu3dspt,senthilashvin,xu2dspt}, but a complete
classification based on this field theory is still demanded.

The goal of this paper is to systematically classify and describe
bosonic SPT phases with various continuous and discrete symmetries
in {\it all dimensions}, using semiclassical nonlinear Sigma model
(NLSM) field theories. At least in one dimensional systems,
semiclassical NLSMs have been proved successful in describing SPT
phases. The O(3) NLSM plus a topological $\Theta-$term describes a
spin-1 Heisenberg chain when $\Theta = 2\pi$: \beqn
\mathcal{S}_{1d} = \int dx d\tau \frac{1}{g} (\partial_\mu
\vec{n})^2 + \frac{i 2\pi}{8\pi} \epsilon_{abc}\epsilon_{\mu\nu}
n^a \partial_\mu n^b
\partial_\nu n^c, \label{o3theta}\eeqn and it is well-known that
the spin-1 antiferromagnetic Heisenberg model is a SPT phase with
2-fold degeneracy at each
boundary~\cite{haldane1,haldane2,affleck1987,kennedy1990,hagiwara1990,ng1994}.

In this paper we will discuss SPT phases with symmetry $Z_2^T$,
$Z_2$, $Z_2 \times Z_2$, $Z_2 \times Z_2^T$, $U(1)$, $U(1) \times
Z_2$, $U(1) \rtimes Z_2$, $U(1) \times Z_2^T$, $U(1) \rtimes
Z_2^T$, $Z_m$, $Z_m \times Z_2$, $Z_m \rtimes Z_2$, $Z_m \times
Z_2^T$, $Z_m \rtimes Z_2^T$, $SO(3)$, $SO(3) \times Z_2^T$, $Z_2
\times Z_2 \times Z_2$. Here we use the standard notation: $Z_2^T$
stands for time-reversal symmetry, $G \times Z_2^T$ and $G \rtimes
Z_2^T$ stand for direct and semidirect product between unitary
group $G$ and time-reversal symmetry. A semidirect product between
two groups means that these two group actions do not commute with
each other. More details will be explained when we discuss the
classification of these states. We will demonstrate that a
$d-$dimensional SPT phase with any symmetry mentioned above can
always be described by an O($d+2$) NLSM in $(d+1)-$dimensional
space-time, namely all the 1d SPT phases discussed in this paper
can be described by Eq.~\ref{o3theta}, all the 2d and 3d SPT
phases can be described by the following two field theories: \beqn
\mathcal{S}_{2d} &=& \int d^2x d\tau \ \frac{1}{g} (\partial_\mu
\vec{n})^2 \cr\cr &+& \frac{i 2\pi k}{\Omega_3} \epsilon_{abcd}
n^a
\partial_\tau n^b
\partial_x n^c \partial_y n^d, \label{o4theta} \eeqn \beqn
\mathcal{S}_{3d} &=& \int d^3x d\tau \ \frac{1}{g} (\partial_\mu
\vec{n})^2 \cr\cr &+& \frac{i 2\pi}{\Omega_4} \epsilon_{abcde} n^a
\partial_\tau n^b
\partial_x n^c \partial_y n^d \partial_z n^e, \label{o5theta}\eeqn
The O($d+2$) vector is a Landau order parameter with a unit length
constraint: $(\vec{n})^2 = 1$. $\Omega_d$ is the surface area of a
$d-$dimensional unit sphere. The $2d$ action Eq.~\ref{o4theta} has
a level$-k$ in front of its $\Theta-$term, whose reason will be
explained later. Different SPT phases in the same dimension are
distinguished by the transformation of the O($d+2$) vector under
the symmetry. The classification of SPT phases on a
$d-$dimensional lattice is given by all the {\it independent}
symmetry transformations of $\vec{n}$ that keep the entire
Lagrangian (including the $\Theta-$term) invariant. This
classification rule will be further clarified in the next section.

An O($d+2$) NLSM can support maximally O($d+2$) symmetry and other
discrete symmetries such as time-reversal. We choose the 17
symmetries listed above, because they can all be embedded into the
maximal symmetry of the field theory, and they are the most
physically relevant symmetries. Of course, if we want to study an
SPT phase with a large Lie group such as SU(N), the above field
theories need to be generalized to NLSM defined with a symmetric
space of that Lie group. But for all these physically relevant
symmetries, our NLSM is already sufficient.

In principle, a NLSM describes a system with a long correlation
length. Thus a NLSM plus a $\Theta-$term most precisely describes
a SPT phase tuned {\it close to} a critical point (but still in
the SPT phase). When a SPT phase is tuned close to a critical
point, the NLSM not only describes its topological properties
($e.g.$ edge states $etc.$), but also describes its dynamics, for
example excitation spectrum above the energy gap (much smaller
than the ultraviolet cut-off). When the system is tuned deep
inside the SPT phase, namely the correlation length is comparable
with the lattice constant, this NLSM can no longer describe its
dynamics accurately, but since the topological properties of this
SPT phase is unchanged while tuning, these topological properties
(like edge states) can still be described by the NLSM. The NLSM is
an effective method of describing the universal topological
properties, as long as we ignore the extra nonuniversal
information about dynamics, such as the exact dispersion of
excitations, which depends on the details of the lattice
Hamiltonian and hence is not universal.

Besides the classification, our NLSMs in all dimensions can tell
us explicit physical information about this SPT phase. For
example, the boundary states of 1d SPT phases can be obtained by
explicitly solving the field theory reduced to the 0d boundary.
The boundary of a 3d SPT phase could be a 2d topological phase,
and the NLSMs can tell us the quantum number of the anyons of the
boundary topological phases. The boundary topological phases of 3d
SPT phases with $U(1)$ and time-reversal symmetry were discussed
in Ref.~\onlinecite{senthilashvin}. We will analyze the boundary
topological phases for some other 3d SPT phases in the current
paper.

Our formalism not only can study each individual SPT phase, it
also reveals the relation between different SPT phases. For
example, using our formalism we are able to show that there is a
very intriguing relation between SPT phases with $U(1)\times
(\rtimes) G$ symmetry and SPT phases with $Z_m \times (\rtimes) G$
symmetry, where $G$ is another discrete group such as $Z_2$,
$Z_2^T$. Our formalism demonstrates that after breaking U(1) to
$Z_m$, whether the SPT phase survives or not depends on the parity
of integer $m$. We also demonstrate that when $m$ is an even
number, we can construct some extra SPT phases with $Z_m \times
(\rtimes) G$ symmetry that {\it cannot} be deduced from SPT phases
with $U(1)\times (\rtimes) G$ symmetry by breaking U(1) down to
$Z_m$. Our field theory also gives many of these SPT states a
natural ``decorated defect" construction, which will be discussed
in more detail in the next section.

NLSMs with a $\Theta-$term can also give us the illustrative
universal bulk ground state wave function of the SPT phases. This
was discussed in Ref.~\onlinecite{xusenthil}. These wave functions
contain important information for both the boundary and the bulk
defects introduced by coupling the NLSM to an external gauge
field~\cite{xusenthil,xulinedefect}. It was also demonstrated that
the NLSMs are useful in classifying and describing symmetry
enriched topological (SET) phases~\cite{xuset}, but a complete
classification of SET phases based on NLSMs will be studied in the
future.

In the current paper we will only discuss SPT states within
cohomology. It is now understood that the group cohomology
classification is incomplete, and in each dimension there are a
few examples beyond cohomology
classification~\cite{kapustin1,kapustin4,kongwen}. These
beyond-cohomology states all involve gravitational
anomalies~\cite{xu6d} or mixed gauge-gravitational
anomalies~\cite{kongwen}. Generalization of our field theory to
the cases beyond group cohomology can be found in another
paper~\cite{xubeyond}.

\section{Strategy and Clarification}

\subsection{Edge states of NLSMs with $\Theta-$term}

In $d-$dimensional theories Eq.~\ref{o3theta},\ref{o4theta} and
\ref{o5theta} ($d$ denotes the spatial dimension), when $\Theta =
2\pi$, their boundaries are described by $(d-1)+1-$dimensional
O($d+2$) NLSMs with a Wess-Zumino-Witten (WZW) term at level-1.
When $d = 1$, the boundary of Eq.~\ref{o3theta} with $\Theta =
2\pi$ is a 0+1d O(3) NLSM with a Wess-Zumino-Witten term at level
$ k = 1 $~\cite{ng1994}: \beqn \mathcal{S}_b = \int d\tau
\frac{1}{g} (\partial_\tau \vec{n})^2 + \int d\tau du \frac{i
2\pi}{8\pi} \epsilon_{abc}\epsilon_{\mu\nu} n^a
\partial_\mu n^b \partial_\nu n^c. \eeqn The WZW term involves an extension of
$\vec{n}(\tau)$ to $\vec{n}(\tau, u)$: \beqn \vec{n}(\tau, 0) =
(0, 0, 1), \ \ \ \vec{n}(\tau, 1) = \vec{n}(\tau). \eeqn The
boundary action $\mathcal{S}_b$ describes a point particle moving
on a sphere $S^2$, with a $2\pi$ magnetic flux through the sphere.
The ground state of this single particle quantum mechanics problem
is two fold degenerate. The two fold degenerate ground states have
the following wave functions on the unit sphere: \beqn && U =
(\cos(\theta/2)e^{i\phi/2}, \ \ \sin(\theta/2)e^{- i\phi/2})^t,
\cr\cr && \vec{n} = \left(\sin(\theta)\cos(\phi),
\sin(\theta)\sin(\phi), \cos(\theta) \right). \label{o3wf}\eeqn
The boundary doublet $U$ transforms projectively under symmetry of
the SPT phase, and its transformation can be derived explicitly
from the transformation of $\vec{n}$. For example if $\vec{n}$
transforms as $\vec{n} \rightarrow -\vec{n}$ under time-reversal,
then this implies that under time-reversal $\phi \rightarrow
\phi$, $\theta \rightarrow \pi + \theta$, and $U \rightarrow i
\sigma^y U$.

When $d = 2$, the boundary is a 1+1-dimensional O(4) NLSM with a
WZW term at level $k = 1$, and it is well-known that this theory
is a gapless conformal field theory if the system has a full O(4)
symmetry~\cite{witten1984,KnizhnikZamolodchikov1984}. The 1d
boundary could be gapped but still degenerate if the symmetry of
$\vec{n}$ is discrete (the degeneracy corresponds to spontaneous
discrete symmetry breaking); when $d = 3$, the boundary is a 2+1d
O(5) NLSM with a WZW at level $k = 1$, which can be reduced to a
2+1d O(4) NLSM with $\Theta = \pi$ after the fifth component of
$\vec{n}$ is integrated out~\cite{senthilashvin}. This $2+1d$
boundary theory should either be gapless or degenerate, and one
particularly interesting possibility is that it can become a
topological order, which will be discussed in more detail in
section IIF. Starting with this topological order, we can prove
that this $2+1d$ boundary system cannot be gapped without
degeneracy.

All components of $\vec{n}$ in Eq.~\ref{o3theta},\ref{o4theta} and
\ref{o5theta} must have a nontrivial transformation under the
symmetry group $G$, namely it is not allowed to turn on a linear
``Zeeman" term that polarizes any component of $\vec{n}$.
Otherwise the edge states can be trivially gapped, and the bulk
$\Theta-$term plays no role.

\subsection{Phase diagram of NLSMs with a $\Theta-$term }

In our classification, the NLSM including its $\Theta-$term is
invariant under the symmetry of the SPT phase, for arbitrary value
of $\Theta$. For special values of $\Theta$, such as $\Theta = k
\pi$ with integer $k$, some extra discrete symmetry may emerge,
but these symmetries are {\it unimportant} to the SPT phase.
However, these extra symmetries guarantee that $\Theta = k \pi$ is
a fixed point under renormalization group (RG) flow. In 1+1d
NLSMs, the RG flow of $\Theta$ was calculated explicitly in
Ref.~\onlinecite{pruisken1,pruisken2} and it was shown that
$\Theta = 2\pi k$ are stable fixed points, while $\Theta =
(2k+1)\pi$ are instable fixed points, which correspond to phase
transitions; in higher dimensions, similar explicit calculations
are possible, but for our purposes, we just need to argue that
$\Theta = 2\pi k$ are stable fixed points under RG flow. The bulk
spectrum of the NLSM with $\Theta = 2\pi k$ is identical to the
case with $\Theta = 0$: in the quantum disordered phase the bulk
of the system is fully gapped without degeneracy. Now if $\Theta$
is tuned away from $2\pi k$: $\Theta = 2\pi k \pm \epsilon$, this
perturbation cannot close the bulk gap, and since the essential
symmetry of the SPT phase is unchanged, the SPT phase including
its edge states should be stable against this perturbation. Thus a
SPT phase corresponds to a finite phase $\Theta \in (2\pi k -
\delta_1, 2\pi k + \delta_2)$ in the phase diagram.

There is a major difference between $\Theta-$term in NLSM and the
$\Theta-$term in the response action of the external gauge field.
In our description, a SPT phase corresponds to the entire phase
whose stable fixed point is at $\Theta = 2\pi$ (or $2\pi k$ with
integer $k$). Tuning slightly away from these stable fixed points
will not break any essential symmetry that protects the SPT state,
and hence it does not change the main physics. The theory will
always flow back to these stable fixed points under RG (this RG
flow was computed explicitly in $1+1d$ in
Ref.~\onlinecite{pruisken1,pruisken2}, and a similar RG flow was
proposed for higher dimensional cases~\cite{xuludwig2013}). The
$\Theta-$term of the external gauge field after integrating out
the matter fields is protected by the symmetry of the SPT phase to
be certain discrete value. For example $\Theta = \pi$ for the
ordinary 3d topological insulator~\cite{qi2008,mooretheta} is
protected by time-reversal symmetry. Tuning $\Theta$ away from
$\pi$ will necessarily break the time-reversal symmetry.


\subsection{$\mathbb{Z}_k$ or $\mathbb{Z}$ classification?}

In the classification table in Ref.~\onlinecite{wenspt,wenspt2},
one can see that in even dimensions, there are many SPT states
with $\mathbb{Z}$ classifications, but in odd dimensions,
$\mathbb{Z}$ classification never appears. This fact was a
consequence of mathematical calculations in
Ref.~\onlinecite{wenspt,wenspt2}, but in this section we will give
a very simple explanation based on our field theories.

The manifold of O($d+2$) NLSM is S$^{d+1}$, which has a
$\Theta-$term in $(d+1)-$dimensional space-time due to homotopy
group $\pi_{d+1}[S^{d+1}] = \mathbb{Z}$. However, this does {\it
not} mean that the $\Theta-$term will always give us $\mathbb{Z}$
classification, because more often than not we can show that
$\Theta = 0$ and $\Theta = 2\pi k$ with certain nonzero integer
$k$ can be connected to each other without any bulk transition.

For example, let us couple two Haldane phases to each other: \beqn
\mathcal{L} &=& \frac{1}{g} (\partial_\mu \vec{n}^{(1)})^2 +
\frac{i 2\pi}{8\pi}\epsilon_{abc}\epsilon_{\mu\nu} n_a^{(1)}
\partial_\mu n_b^{(1)} \partial_\nu n_c^{(1)} \cr\cr &+& 1 \rightarrow 2 +
A (\vec{n}^{(1)} \cdot \vec{n}^{(2)}). \label{2o3theta} \eeqn When
$A < 0$, effectively $\vec{n}^{(1)} = \vec{n}^{(2)} = \vec{n}$,
then the system is effectively described by one O(3) NLSM with
$\Theta = 4\pi$; while when $A >0 $, effectively $\vec{n}^{(1)} =
- \vec{n}^{(2)} = \vec{n}$, the effective NLSM for the system has
$\Theta = 0$. When parameter $A$ is tuned from negative to
positive, the bulk gap does not close. The reason is that, since
$\Theta = 2\pi$ in both Haldane phases, the $\Theta-$term does not
affect the bulk spectrum at all. To analyze the bulk spectrum (and
bulk phase transition) while tuning $A$, we can just ignore the
$\Theta-$term. Without the $ \Theta-$term, both theories are just
trivial gapped phases, and an inter-chain coupling can not
qualitatively change the bulk spectrum unless it is strong enough
to overcome the bulk gap in each chain. We have explicitly checked
this phase diagram using a Monte Carlo simulation of two coupled
O(3) NLSMs, and the result is exactly the same as what we would
expect from the argument above. Thus the theory with $\Theta =
4\pi$ and $\Theta = 0$ are equivalent.

By contrast, if we couple two chains with $\Theta = \pi$ each,
then the cases $A > 0$ and $< 0$ correspond to effective $\Theta =
0$ and $2\pi$ respectively, and these two limits are separated by
a bulk phase transition point $A = 0$, when the system becomes two
decoupled chains with $\Theta = \pi$ each. And it is well-known
that a $1+1d$ O(3) NLSM with $\Theta = \pi$ is the effective field
theory that describes a spin-1/2 chain~\cite{haldane1,haldane2},
and according to the Lieb-Shultz-Matthis theorem, this theory must
be either gapless or degenerate~\cite{LSM}. This conclusion is
consistent with the RG calculation in
Ref.~\onlinecite{pruisken1,pruisken2}, and a general
nonperturbative argument in Ref.~\onlinecite{xuludwig2013}.

In fact when $\Theta = 4\pi$ the boundary state of
Eq.~\ref{o3theta} is a spin-1 triplet, and by tuning $A$, at the
boundary there is a level crossing between triplet and singlet,
while there is no bulk transition. This analysis implies that with
SO(3) symmetry, $1d$ spin systems have two different classes:
there is a trivial class with $\Theta = 4\pi k$, and a nontrivial
Haldane class with $\Theta = (4k + 2)\pi$.

If we cannot connect $\Theta = 4\pi$ to $\Theta = 0$ without
closing the bulk gap, then the classification would be bigger than
$\mathbb{Z}_2$. For example, let us consider the 2d SPT phase with
U(1) symmetry which was first studied in
Ref.~\onlinecite{levinsenthil}. This phase is described by
Eq.~\ref{o4theta}. $B \sim n_1 + i n_2$ and $B^\prime \sim n_3 + i
n_4$ ($n_1 \cdots n_4$ are the four components of O(4) vector
$\vec{n}$ in Eq.~\ref{o4theta}) are two complex boson (rotor)
fields that transform identically under the global U(1) symmetry.
Now suppose we couple two copies of this systems together through
symmetry allowed interactions: \beqn \mathcal{S} &=& \mathcal{S}_1
+ \mathcal{S}_2 + A_1 B_1 B_2^\dagger + A_2 B_1 B_2^{\prime
\dagger} \cr\cr &+& A_3B^\prime_1 B_2^{\dagger} + A_4 B^\prime_1
B_2^{\prime \dagger} + H.c. \eeqn No matter how we tune the
parameters $A_i$, the resulting effective NLSM {\it always} has
$\Theta = 4\pi$ instead of $\Theta = 0$ (this is simply because
$(-1)^2 = (-1)^4 = +1$). This implies that we cannot smoothly
connect $\Theta = 4\pi$ to $0$ without any bulk transition. Thus
the classification of 2d SPT phases with U(1) symmetry is
$\mathbb{Z}$ instead of $\mathbb{Z}_2$. This is why in $2d$ (and
all even dimensions), many SPT states have $\mathbb{Z}$
classification, while in odd dimensions there is no $\mathbb{Z}$
classification at all, namely all the nontrivial SPT phases in odd
dimensions correspond to $\Theta = 2\pi$. Thus in
Eq.~\ref{o4theta} we added a level$-k$ in the $\Theta-$term.

\subsection{NLSM and ``decorated defect" construction of SPT states}

Ref.~\onlinecite{chenluashvin} has given us a physical
construction of some of the SPT states in terms of the ``decorated
domain wall" picture. For example, one of the $3d$ $Z^A_2 \times
Z^B_2$ SPT state can be constructed as follows: we first break the
$Z_2^B$ symmetry, then restore the $Z_2^B$ symmetry by
proliferating the domain wall of $Z_2^B$, and each $Z_2^B$ domain
wall is decorated with a $2d$ SPT state with $Z_2^A$ symmetry.
This state is described by Eq.~\ref{o5theta} with transformation
\beqn Z_2^B &:& n_{1, 2} \rightarrow - n_{1, 2}, \ \ n_a
\rightarrow n_a (a = 3,4,5); \cr\cr Z_2^A &:& n_1,\rightarrow n_1,
\ \ n_a \rightarrow - n_a (a = 2, \cdots 5). \eeqn Here $n_i$ is
the $i$th component of vector $\vec{n}$. To visualize the
``decorated domain" wall picture, we can literally make a domain
wall of $n_1$, and consider the following configuration of vector
$\vec{n}$: $\vec{n} = (\cos\theta, \sin\theta N_2, \sin\theta N_3,
\sin\theta N_4, \sin\theta N_5)$, where $\vec{N}$ is a O(4) vector
with unit length, and $\theta$ is a function of coordinate $z$
only: \beqn \theta(z = +\infty) = \pi, \ \ \ \theta(z = - \infty)
= 0. \eeqn Plug this parametrization of $\vec{n}$ into
Eq.~\ref{o5theta}, and integrate along $z$ direction, the
$\Theta-$term in Eq.~\ref{o5theta} precisely reduces to the
$\Theta-$term in Eq.~\ref{o4theta} with $k = 1$, and the O(4)
vector $\vec{n} = \vec{N}$. This is precisely the $2d$ SPT with
$Z_2$ symmetry. This implies that the $Z_2^B$ domain wall is
decorated with a $2d$ SPT state with $Z_2^A$ symmetry.

Many SPT states can be constructed with this decorated domain wall
picture. Some $3d$ SPT states can also be understood as
``decorated vortex", which was first discussed
in~\onlinecite{senthilashvin}. This state has $U(1) \times Z_2^T$
symmetry, and the vector $\vec{n}$ transforms as \beqn U(1) &:&
(n_1 + in_2) \rightarrow (n_1 + in_2)e^{i\theta}, \ \ n_{3,4,5}
\rightarrow n_{3,4,5}, \cr\cr Z_2^T &:& \vec{n} \rightarrow
-\vec{n}. \eeqn If we make a vortex of the U(1) order parameter
$(n_1, n_2)$, Eq.~\ref{o5theta} reduces to Eq.~\ref{o3theta} with
O(3) order parameter $(n_3, n_4, n_5)$. Thus this SPT can be
viewed as decorating the U(1) vortex with a $1d$ Haldane phase,
and then proliferating the vortices.

\subsection{Independent NLSMs}

Let us take the example of 1d SPT phases with $Z_2 \times Z_2^T$
symmetry. As we claimed, all 1d SPT phases in this paper are
described by the same NLSM Eq.~\ref{o3theta}. With $Z_2 \times
Z_2^T$ symmetry, there seems to be three different ways of
assigning transformations to $\vec{n}$ that make the entire
Lagrangian invariant: \beqn (1)&:& Z_2: \vec{n} \rightarrow
\vec{n}, \ \ Z_2^T: \vec{n} \rightarrow - \vec{n}. \cr\cr (2)&:&
Z_2: n_{1,2} \rightarrow - n_{1,2}, \ \ n_3 \rightarrow n_3 \cr
\cr && Z_2^T: \vec{n} \rightarrow - \vec{n} \cr\cr (3) &:& Z_2:
n_{1,2} \rightarrow - n_{1,2}, \ \ n_3 \rightarrow n_3 \cr \cr &&
Z_2^T: n_3 \rightarrow -n_3, \ \ n_{1,2} \rightarrow n_{1,2}.
\label{option}\eeqn However the NLSMs defined with these three
different transformations are not totally independent from each
other. Let us parameterize the O(3) vectors $\vec{n}^{(i)}$ with
transformations (1), (2) and (3) as: \beqn \vec{n}^{(i)}(\vec{r})
= (n^{(i)}_1, n^{(i)}_2, n^{(i)}_3 ) = \cr\cr \left(
\sin(\theta^{(i)}_{\vec{r}})\cos(\phi^{(i)}_{\vec{r}}),
\sin(\theta^{(i)}_{\vec{r}})\sin(\phi^{(i)}_{\vec{r}}),
\cos(\theta^{(i)}_{\vec{r}}) \right), \eeqn $\phi^{(i)}_{\vec{r}}$
and $\theta^{(i)}_{\vec{r}}$ are functions of space-time. Under
$Z_2$ and $Z_2^T$ symmetry, $\theta^{(i)}$ and $\phi^{(i)}$
transform as \beqn Z_2 &:& \theta^{(i)} \rightarrow \theta^{(i)},
\cr\cr && \phi^{(1)} \rightarrow \phi^{(1)}, \ \ \phi^{(i)}
\rightarrow \phi^{(i)} + \pi, \ (i = 2, 3); \cr\cr Z_2^T &:&
\theta^{(i)} \rightarrow \pi - \theta^{(i)}, \cr\cr && \phi^{(i)}
\rightarrow \phi^{(i)} + \pi, \ (i = 1,2), \ \ \phi^{(3)}
\rightarrow \phi^{(3)}. \eeqn First of all, since $\theta^{(i)}$
have the same transformation for all $i$, we can turn on strong
coupling between the three NLSMs to make $\theta^{(1)} =
\theta^{(2)} = \theta^{(3)} = \theta$. Now we can construct
$\vec{n}^{(3)}$ using the parametrization of $\vec{n}^{(1)}$ and
$\vec{n}^{(2)}$: \beqn n^{(3)}_1 &=& \sin(\theta) \cos(\phi^{(1)}
+ \phi^{(2)}), \cr\cr n^{(3)}_2 &=& \sin(\theta) \sin(\phi^{(1)} +
\phi^{(2)}), \cr\cr n^{(3)}_3 &=& \cos(\theta). \eeqn It is
straightforward to prove that $\vec{n}^{(3)}$ defined this way
transforms identically with the case $(3)$ in Eq.~\ref{option},
also the topological number of $\vec{n}^{(3)}$ in 1+1d space-time
is the sum of topological numbers of $\vec{n}^{(1)}$ and
$\vec{n}^{(2)}$. More explicitly, an instanton of $\vec{n}^{(a)}$
is a domain wall of $n^{(a)}_3$ decorated with a vortex of
$\phi^{(a)}$. As we explained above, with appropriate coupling
between these vectors, we can make $\theta^{(1)} = \theta^{(2)} =
\theta^{(3)} = \theta$, and $\phi^{(3)} = \phi^{(1)} +
\phi^{(2)}$. Thus a domain wall of $n^{(3)}_3$ is also a domain
wall of $n^{(1)}_3$ and $n^{(2)}_3$, while the vortex number of
$\phi^{(3)}$ is the sum of vortex number of $\phi^{(1)}$ and
$\phi^{(2)}$. Thus the $\Theta-$term of $\vec{n}^{(3)}$ reduces to
the sum of $\Theta-$terms of $\vec{n}^{(1)}$ and $\vec{n}^{(2)}$.
In this example we have shown that NLSMs (1) and (2) in
Eq.~\ref{option} can ``merge" into NLSM (3). Thus the three NLSMs
defined with transformations (1), (2) and (3) are not independent
from each other.~\footnote{The ``merging" argument is usually easy
to implement for systems with simple symmetries, but we should
admit that for higher dimensions and complicated symmetries, the
``merging" argument can become rather involved.} The consequence
of this analysis is that if all three theories exist in one
system, although each theory is a nontrivial SPT phase
individually, we can turn on some symmetry allowed couplings
between these NLSMs and cancel the bulk topological terms
completely, and drive the entire coupled system to a trivial
state.

Also, for either NLSM $(1)$ or $(2)$ in Eq.~\ref{option}, we can
show that $\Theta^{(i)} = 0$ and $4\pi$ can be connected to each
other without a bulk transition (using the same method as the
previous subsection). Then eventually the 1d SPT phase with $Z_2
\times Z_2^T$ symmetry is parametrized by two independent
$\Theta-$terms, the fixed point values of $\Theta^{(1)}$ and
$\Theta^{(2)}$ can be either $0$ or $2\pi$, thus this SPT phase
has a $(\mathbb{Z}_2)^2$ classification, which is consistent with
the classification using group cohomology. NLSMs with
transformations $(1)$, $(2)$ are two ``root phases" of 1d SPT
phases with $Z_2 \times Z_2^T$ symmetry. All the other SPT phases
can be constructed with these two root phases.

For most SPT phases, we can construct the NLSMs using the smallest
representation (fundamental representation) of the symmetry groups
$G$, because usually (but not always!) NLSMs constructed using
higher representations can reduce to constructions with the
fundamental representation with a different $\Theta$. For example,
the 1d SPT phase with $U(1)\rtimes Z_2$ symmetry can be described
by Eq.~\ref{o3theta} with the following transformation \beqn
U(1)&:& (n_1 + in_2) \rightarrow e^{i\theta} (n_1 + in_2), \ \ n_3
\rightarrow n_3, \cr\cr Z_2 &:& n_1 \rightarrow n_1, \ \ n_{2,3}
\rightarrow - n_{2,3}, \eeqn namely $B \sim (n_1 + in_2)$ is a
charge-1 boson under the U(1) rotation, and the edge state of this
SPT phase carries charge-1/2 of boson $B$. We can also construct
an O(3) NLSM using charge-2 boson $B^\prime \sim (n_1^\prime +
in_2^\prime) \sim (n_1 + in_2)^2$ that transforms as $B^\prime
\rightarrow B^\prime e^{2i \alpha}$, then mathematically we can
demonstrate that the NLSM with $\Theta = 2\pi$ for order parameter
$\vec{n}^\prime = (n_1^\prime, n_2^\prime, n_3)$ reduces to a NLSM
of $\vec{n}$ with $\Theta = 4\pi$, hence it is a trivial phase.

More explicitly, let us take unit vector $\vec{n} = \left(
\sin(\theta)\cos(\phi), \sin(\theta)\sin(\phi), \cos(\theta)
\right)$, and vector $\vec{n}^\prime = \left(
\sin(\theta)\cos(2\phi), \sin(\theta)\sin(2\phi), \cos(\theta)
\right)$, then we can show that when $\vec{n}$ has topological
number 1 in 1+1d space-time, $\vec{n}^\prime$ would have
topological number 2. This means that if there is a $\Theta-$term
for $\vec{n}^\prime$ with $\Theta = 2\pi$, it is equivalent to a
$\Theta-$term for $\vec{n}$ with $\Theta = 4\pi$.

Physically, the edge state of NLSM of $\vec{n}^\prime$ with
$\Theta = 2\pi$ carries a half-charge of $B^\prime$, which is
still a charge-1 object, so it can be screened by another charge-1
boson $B$. Hence in this case NLSM constructed using charge-2
boson $B^\prime$ would be trivial.

However, later we will also show that when the symmetry group
involves $Z_m$ with even integer $m > 2$, then using higher
representations of $Z_m$ we can construct SPT phases that {\it
cannot} be obtained from the fundamental representation of $Z_m$.

\subsection{Boundary topological order of 3d SPT phases}

The $(d-1)-$dimensional boundary of a $d-$dimensional SPT phase
must be either degenerate or gapless. When $d = 3$, its 2d
boundary can spontaneously break the symmetry, or have a
topological order~\cite{senthilashvin}. We can use the bulk field
theory Eq.~\ref{o5theta} to derive the quantum numbers of the
anyons at the boundary.

Let us take the 3d SPT phase with $Z_2 \times Z_2^T$ symmetry as
an example. One of the SPT phases has the following
transformations: \beqn Z_2 &:& n_{a} \rightarrow - n_{a} (a = 1,
\cdots 4), \ \ \ n_5 \rightarrow n_5 ; \cr\cr Z_2^T &:& \vec{n}
\rightarrow -\vec{n}. \eeqn The 2+1d boundary of the system is
described by a 2+1d O(5) NLSM with a Wess-Zumino-Witten (WZW) term
at level $k = 1$: \beqn S &=& \int d^2x d\tau \ \frac{1}{g}
(\partial_\mu \vec{n})^2 \cr\cr &+& \int_0^1 du \ \frac{i 2\pi}{
\Omega_4 } \epsilon_{abcde} n^a
\partial_x n^b \partial_y n^c
\partial_z n^d \partial_\tau n^e, \label{2do5nlsm} \eeqn where $\vec{n}(x, \tau,
u)$ satisfies $\vec{n}(x, \tau, 0) = (0,0,0,0,1)$ and $\vec{n}(x,
\tau, 1) = \vec{n}(x, \tau)$. If the time-reversal symmetry is
preserved, namely $\langle n_5 \rangle = 0$, we can integrate out
$n_5$, and Eq.~\ref{2do5nlsm} reduces to a 2+1d O(4) NLSM with
$\Theta = \pi$: \beqn S = \int d^2x d\tau \ \frac{1}{g}
(\partial_\mu \vec{n})^2 + \frac{i \pi}{ \Omega_3 }
\epsilon_{abcd} n^a
\partial_\tau n^b \partial_x n^c
\partial_y n^d. \label{2do4} \eeqn In Eq.~\ref{2do4} $\Theta =
\pi$ is protected by time-reversal symmetry.

In the following we will argue that the topological terms in
Eq.~\ref{2do5nlsm} and Eq.~\ref{2do4} guarantee that the $2d$
boundary cannot be gapped without degeneracy. One particularly
interesting possibility of the boundary is a phase with 2d $Z_2$
topological order~\cite{senthilashvin}. A 2d $Z_2$ topological
phase has $e$ and $m$ excitations that have mutual semion
statistics~\cite{kitaev2003}. The semion statistics can be
directly read off from Eq.~\ref{2do4}: if we define complex boson
fields $z_1 = n_1 + in_2$ and $z_2 = n_3 + in_4$, then the
$\Theta-$term in Eq.~\ref{2do4} implies that a vortex of $(n_3,
n_4)$ carries half charge of $z_1$, while a vortex of $(n_1, n_2)$
carries half charge of $z_2$, thus vortices of $z_1$ and $z_2$ are
bosons with mutual semion statistics. This statistics survives
after $z_1$ and $z_2$ are disordered by condensing the {\it double
vortex} (vortex with vorticity $4\pi$) of either $z_1$ or $z_2$ at
the boundary, then the disordered phase must inherit the
statistics and become a $Z_2$ topological
phase~\cite{senthilashvin}. The vortices of $(n_1, n_2)$ and
$(n_3, n_4)$ become the $e$ and $m$ excitations respectively.
Normally a vortex defect is discussed in systems with a U(1)
global symmetry. We do not assume such U(1) global symmetry in our
case, this symmetry reduction is unimportant in the $Z_2$
topological phase.

At the vortex core of $(n_3, n_4)$, namely the $m$ excitation,
Eq.~\ref{2do5nlsm} reduces to a $0+1d$ O(3) NLSM with a WZW term
at level 1~\cite{groversenthil}: \beqn \mathcal{S}_m = \int d\tau
\frac{1}{g} (\partial_\tau \vec{N})^2 + \int_0^1 du \frac{i2\pi
}{8\pi} \epsilon_{abc}\epsilon_{\mu\nu} N^a
\partial_\mu N^b
\partial_\nu N^c, \label{0dwzw}\eeqn where $\vec{N} \sim (n_1, n_2, n_5)$. This
0+1d field theory describes a single particle moving on a 2d
sphere with a magnetic monopole at the origin. It is well known
that if there is a SO(3) symmetry for $\vec{N}$, then the ground
state of this 0d problem has two fold degeneracy, with two
orthogonal solutions \beqn && u_m = \cos(\theta/2)e^{i\phi/2}, \ \
\ v_m = \sin(\theta/2)e^{- i\phi/2}, \cr\cr && \vec{N} =
\left(\sin(\theta)\cos(\phi), \sin(\theta)\sin(\phi), \cos(\theta)
\right). \label{doublet}\eeqn Likewise, the vortex of $(n_1, n_2)$
($e$ excitation) also carries a doublet $(u_e, v_e)$. Under the
$Z_2$ transformation, $\phi \rightarrow \phi + \pi$, thus
$u_{e,m}$ and $v_{e,m}$ carry charge $\pm 1/2$ of the $Z_2$
symmetry, namely under the $Z_2$ transformation: \beqn Z_2:
U_{e,m} \rightarrow i\sigma^z U_{e,m}, \eeqn where $U_{e,m} =
(u_{e,m}, v_{e,m})^t$.

Under time-reversal transformation $\mathcal{T}$, $\vec{N}
\rightarrow - \vec{N}$, $\theta \rightarrow \theta + \pi$. Thus
the $e$ and $m$ doublets transform as \beqn Z_2^T: U_{e,m}
\rightarrow i\sigma^y U_{e,m}, \eeqn thus the $e$ and $m$ anyons
at the boundary carry projective representation of $Z_2^T$ which
satisfies $\mathcal{T}^2 = -1$.

Based on this $Z_2$ topological order, we can derive the phase
diagram around the $Z_2$ topological order, and show that this
boundary cannot be gapped without degeneracy. For example,
starting with a $2d$ $Z_2$ topological order, one can condense
either $e$ or $m$ excitation and kill the topological degeneracy.
However, because $U_{e,m}$ transform nontrivially under the
symmetry group, condensate of either $e$ or $m$ will always
spontaneously break certain symmetry and lead to degeneracy. For
example, the condensate of $e$ excitation has nonzero expectation
value of $(n_3, n_4, n_5) \sim U^\dagger_e \vec{\sigma} U_e$,
which necessarily spontaneously breaks the $Z_2$ or $Z_2^T$
symmetry.

We also note that one bulk BSPT state can have different boundary
states, which depends on the details of the boundary Hamiltonian.
Recently a different boundary topological order of BSPT state was
derived in Ref.~\onlinecite{lukaszsemion}, but the bulk state is
the same as ours.

\subsection{Rule of classification}

With all these preparations, we are ready to lay out the rules of
our classification:

{\it 1.} In $d-$dimensional space, all the SPT phases discussed in
this paper are described by a $(d+1)-$dimensional O($d+2$) NLSM
with a $\Theta-$term. The O($d+2$) vector field $\vec{n}$ is an
order parameter, namely it must carry a nontrivial representation
of the given symmetry. In other words, no component of the vector
field transforms completely trivially under the symmetry, because
otherwise it is allowed to turn on a strong linear ``Zeeman" term
to the trivial component, and then the system will become a
trivial direct product state.

{\it 2.} The classification is given by all the possible {\it
independent} symmetry transformations on vector order parameter
$\vec{n}$ that keep the $\Theta-$term invariant, for {\it
arbitrary} value of $\Theta$. Independent transformations mean
that any NLSM defined with one transformation cannot be obtained
by ``merging" two (or more) other NLSMs defined with other
transformations. SPT phases constructed using independent NLSMs
are called ``root phases". All the other SPT phases can be
constructed with these root phases.

{\it 3.} With a given symmetry, and given transformation of
$\vec{n}$, if $\Theta = 2\pi k$ and $\Theta = 0$ can be connected
without a bulk transition, this transformation will contribute
classification $\mathbb{Z}_k$; otherwise the transformation will
contribute classification $\mathbb{Z}$.

Using the rule and strategy discussed in this section, we can
obtain the classification of all SPT phases in all dimensions. In
this paper we will systematically study SPT phases in one, two and
three spatial dimensions with symmetries $Z_2^T$, $Z_2$, $Z_2
\times Z_2$, $Z_2 \times Z_2^T$, $U(1)$, $U(1) \times Z_2$, $U(1)
\rtimes Z_2$, $U(1) \times Z_2^T$, $U(1) \rtimes Z_2^T$, $Z_m$,
$Z_m \times Z_2$, $Z_m \rtimes Z_2$, $Z_m \times Z_2^T$, $Z_m
\rtimes Z_2^T$, $SO(3)$, $SO(3) \times Z_2^T$, $Z_2 \times Z_2
\times Z_2$. The final classification of the SPT phases we study
in this paper is consistent to the classification based on group
cohomology~\cite{wenspt,wenspt2}.

\section{1d SPT phase with $Z_2 \times Z_2 \times Z_2^T$
symmetry}

Before we discuss our full classification, let us carefully
discuss 1d SPT phases with $Z_2 \times Z_2 \times Z_2^T$ symmetry
as an example. These SPT phases were discussed very thoroughly in
Ref.~\onlinecite{1dd2h}. There are in total 16 different phases
(including the trivial phase). The goal of this section is to show
that all these phases can be described by the {\it same} equation
Eq.~\ref{o3theta} with certain transformation of $\vec{n}$, and
the projective representation of the boundary states given in
Ref.~\onlinecite{1dd2h} can be derived explicitly using
Eq.~\ref{o3wf}.

For the consistency of notation in this paper, $R_z$ and $R_x$ in
Ref.~\onlinecite{1dd2h} will be labelled $Z_2^A$ and $Z_2^B$ here.
Let us consider one example, namely Eq.~\ref{o3theta} with the
following transformation: \beqn Z_2^A &:& n_{1,2} \rightarrow
-n_{1,2}, \ \ n_3 \rightarrow n_3; \cr\cr Z_2^B &:& n_{2,3}
\rightarrow -n_{2,3}, \ \ n_1 \rightarrow n_1; \cr\cr Z_2^T &:&
n_2 \rightarrow - n_2, \ \ n_{1,3} \ra n_{1,3}. \eeqn Now let us
parametrize $\vec{n}$ as \beqn \vec{n} = \left(\sin \theta
\cos\phi, \ \sin\theta \sin\phi, \ \cos\theta \right), \eeqn then
$\theta$ and $\phi$ transform as \beqn Z_2^A &:& \theta \ra
\theta,  \ \ \phi \ra \phi  + \pi, \cr\cr Z_2^B &:& \theta \ra \pi
- \theta, \ \ \phi \ra - \phi, \cr\cr Z_2^T &:& \theta \ra \theta,
\ \ \phi \ra - \phi. \eeqn These transformations lead to the
following projective transformation of edge state Eq.~\ref{o3wf}:
\beqn Z_2^A &:& U \ra i \sigma^z U, \cr\cr Z_2^B &:& U \ra
\sigma^x U, \cr\cr Z_2^T &:& U \ra U. \eeqn Thus this NLSM
corresponds to phase $E_5$ in Ref.~\onlinecite{1dd2h}.

The 16 phases in Ref.~\onlinecite{1dd2h} correspond to the
following transformations of O(3) vector $\vec{n}$: \beqn E_0 &:&
\mathrm{Trivial \ phase,} \ \Theta = 0; \cr\cr E_0^\prime &:&
Z_2^A, Z_2^B : \vec{n} \ra \vec{n}, \ \ Z_2^T : \vec{n} \ra -
\vec{n} ; \cr\cr E_1 &:& Z_2^A : \vec{n} \ra \vec{n}, \cr\cr &&
Z_2^B : n_{1,2} \ra - n_{1,2} , \ \ n_3 \ra n_3, \cr\cr && Z_2^T :
\vec{n} \ra - \vec{n}, \cr\cr E_1^\prime &:& Z_2^A : \vec{n} \ra
\vec{n}, \cr\cr && Z_2^B : n_{1,2} \ra - n_{1,2} , \ \ n_3 \ra
n_3, \cr\cr && Z_2^T : n_{1,2} \ra n_{1,2}, \ \ n_3 \ra - n_3;
\cr\cr E_3 &:& Z_2^B : \vec{n} \ra \vec{n}, \cr\cr && Z_2^A :
n_{1,2} \ra - n_{1,2} , \ \ n_3 \ra n_3, \cr\cr && Z_2^T : \vec{n}
\ra - \vec{n}, \cr\cr E_3^\prime &:& Z_2^B : \vec{n} \ra \vec{n},
\cr\cr && Z_2^A : n_{1,2} \ra - n_{1,2} , \ \ n_3 \ra n_3, \cr\cr
&& Z_2^T : n_{1,2} \ra n_{1,2}, \ \ n_3 \ra - n_3; \cr\cr E_5 &:&
Z_2^A : n_{1,2} \rightarrow -n_{1,2}, \ \ n_3 \rightarrow n_3;
\cr\cr && Z_2^B : n_{2,3} \rightarrow -n_{2,3}, \ \ n_1
\rightarrow n_1; \cr\cr && Z_2^T : n_2 \rightarrow - n_2, \ \
n_{1,3} \ra n_{1,3}; \cr\cr E_5^\prime &:& E_5 \oplus E_0^\prime;
\cr\cr E_7 &:& Z_2^A : n_{1,2} \ra - n_{1,2} , \ \ n_3 \ra n_3,
\cr\cr && Z_2^B : n_{1,2} \ra - n_{1,2} , \ \ n_3 \ra n_3, \cr\cr
&& Z_2^T : n_{1,2} \ra n_{1,2} , \ \ n_3 \ra - n_3; \cr\cr
E_7^\prime &:& Z_2^A : n_{1,2} \ra - n_{1,2} , \ \ n_3 \ra n_3,
\cr\cr && Z_2^B : n_{1,2} \ra - n_{1,2} , \ \ n_3 \ra n_3, \cr\cr
&& Z_2^T : \vec{n} \ra - \vec{n}; \cr\cr E_9 &:& Z_2^A : n_{1,2}
\rightarrow -n_{1,2}, \ \ n_3 \rightarrow n_3; \cr\cr && Z_2^B :
n_{2,3} \rightarrow -n_{2,3}, \ \ n_1 \rightarrow n_1; \cr\cr &&
Z_2^T : n_3 \rightarrow - n_3, \ \ n_{1,2} \ra n_{1,2}; \cr\cr
E_9^\prime &:& E_9 \oplus E_0^\prime, \cr\cr E_{11} &:& Z_2^A :
n_{1,2} \rightarrow -n_{1,2}, \ \ n_3 \rightarrow n_3; \cr\cr &&
Z_2^B : n_{2,3} \rightarrow -n_{2,3}, \ \ n_1 \rightarrow n_1;
\cr\cr && Z_2^T : n_1 \rightarrow - n_1, \ \ n_{2,3} \ra n_{2,3};
\cr\cr E_{11}^\prime &:& E_{11} \oplus E_{0}^\prime ; \cr\cr
E_{13} &:& Z_2^A : n_{1,2} \rightarrow -n_{1,2}, \ \ n_3
\rightarrow n_3; \cr\cr && Z_2^B : n_{2,3} \rightarrow -n_{2,3}, \
\ n_1 \rightarrow n_1; \cr\cr && Z_2^T : \vec{n} \ra -\vec{n};
\cr\cr E_{13}^\prime &:& E_{13} \oplus E_{0}^\prime. \eeqn

All the phases except for the trivial phase $E_0$ have $\Theta =
2\pi$ in Eq.~\ref{o3theta}. Here $E_5 \oplus E_0^\prime$ means it
is a spin ladder with symmetry allowed weak interchain couplings,
and the two chains are $E_5$ phase and $E_0^\prime$ phase
respectively. For all the 16 phases above, we can compute the
projective representations of the boundary states using
Eq.~\ref{o3wf}, and they all precisely match with the results in
Ref.~\onlinecite{1dd2h}.

\section{Full classification of SPT phases}

\subsection{SPT phases with $Z_2$ symmetry}

In 1d and 3d, there is no $Z_2$ symmetry transformation that we
can assign vector $\vec{n}$ that makes the actions
Eq.~\ref{o3theta} and Eq.~\ref{o5theta} invariant, thus there is
no SPT phase in 1d and 3d with $Z_2$ symmetry. However, in 2d
there is obviously one and only one way to assign the $Z_2$
symmetry: \beqn Z_2: (n_1, n_2, n_3, n_4) \rightarrow - (n_1, n_2,
n_3, n_4). \label{2dz2} \eeqn Then when $\Theta = 2\pi$ this 2+1d
O(4) NLSM describes the $Z_2$ SPT phase studied in
Ref.~\onlinecite{levingu}. Using the method in section IIC, one
can show that with the transformation Eq.~\ref{2dz2}, the 2+1d
O(4) NLSM Eq.~\ref{o4theta} with $\Theta = 4\pi$ is equivalent to
$\Theta = 0$, thus the classification in 2d is $\mathbb{Z}_2$.

In Ref.~\onlinecite{xusenthil}, the authors also used this NLSM to
derive the ground state wave function of the SPT phase: \beqn
|\Psi\rangle = \sum (-1)^{dw} |C\rangle, \eeqn where $|C\rangle$
standards for an arbitrary Ising field configuration, while $dw$
is the number of Ising domain walls of this configuration. This
wave function was also derived in Ref.~\onlinecite{levingu} with
an exactly soluble model for this SPT phase.

{\it The classification of SPT phases with $Z_2$ symmetry is}:
\beqn 1d: \mathbb{Z}_1, \ \ \ 2d: \mathbb{Z}_2, \ \ \  3d:
\mathbb{Z}_1. \eeqn Here $\mathbb{Z}_1$ means there is only one
trivial state, and $\mathbb{Z}_2$ means there is one trivial state
and one nontrivial SPT state.

\subsection{SPT phases with $Z_2^T$ symmetry}

In 2d, there is no way to assign $Z_2^T$ symmetry to the O(4) NLSM
order parameter in Eq.~\ref{o4theta} to make the $\Theta-$term
invariant, thus there is no bosonic SPT phase in 2d with $Z_2^T$
symmetry. In 1d and 3d, there is only one way to assign the
$Z_2^T$ symmetry to vector $\vec{n}$: \beqn Z_2^T : \vec{n}
\rightarrow -\vec{n}, \eeqn and $\Theta = 0$ and $\Theta = 4\pi$
are equivalent. Thus in both 1d and 3d, the classification is
$\mathbb{Z}_2$. Notice that time-reversal is an antiunitary
transformation, thus $i \rightarrow -i$ under $Z_2^T$; also since
our NLSMs are defined in Euclidean space-time, the Euclidean time
$\tau = it $ is invariant under $Z_2^T$.

Using the method in section II.F, one can demonstrate that the
boundary of the $3d$ SPT state with $Z_2^T$ symmetry is a $2d$
$Z_2$ topological order, whose both $e$ and $m$ excitations are
Kramers doublet, i.e. the so called $eTmT$ state.

{\it The classification of SPT phases with $Z_2^T$ symmetry is}:
\beqn 1d: \mathbb{Z}_2, \ \ \ 2d: \mathbb{Z}_1, \ \ \ 3d:
\mathbb{Z}_2. \eeqn

Now it is understood that in $3d$ there is bosonic SPT state with
$Z_2^T$ symmetry that is beyond the group cohomology
classification~\cite{senthilashvin}, and there is a explicit
lattice construction for such state~\cite{fionachen}. This state
is also beyond our current NLSM description. However, a
generalized field theory which involves both the NLSM and
Chern-Simons theory can describe at least a large class of BSPT
states beyond group cohomology. This will be discussed in a
different paper~\cite{xubeyond}.

\subsection{SPT phases with $U(1)$ symmetry}

In 1d and 3d, there is no way to assign U(1) symmetry to vector
$\vec{n}$ that keeps the entire Lagrangian invariant. But in 2d,
bosonic SPT phase with U(1) symmetry was discussed in
Ref.~\onlinecite{levinsenthil}, and its field theory is given by
Eq.~\ref{o4theta}. And since in this case we cannot connect
$\Theta = 2\pi k$ and $\Theta = 0$ without a bulk transition, the
classification is $\mathbb{Z}$.

{\it The classification of SPT phases with $U(1)$ symmetry is}:
\beqn 1d: \mathbb{Z}_1, \ \ \ 2d: \mathbb{Z}, \ \ \  3d:
\mathbb{Z}_1. \eeqn

\subsection{SPT phases with $U(1) \rtimes Z_2$ symmetry}

$U(1)\rtimes Z_2$ is a subgroup of SO(3). In 1d, there is only one
way of assigning the symmetry to vector $\vec{n}$ that keeps the
entire Lagrangian invariant: \beqn U(1)&:& (n_1 + in_2)
\rightarrow e^{i\theta} (n_1 + in_2), \ \ n_3 \rightarrow n_3,
\cr\cr Z_2 &:& n_1 \rightarrow n_1, \ \ n_{2,3} \rightarrow -
n_{2,3}. \label{1du1rz2}\eeqn Here $Z_2$ is a particle-hole
transformation of rotor/boson field $b \sim n_1 + in_2$. $n_3$ can
be viewed as the boson density, which changes sign under
particle-hole transformation. One can check that the U(1) and
$Z_2$ symmetry defined above do not commute with each other. The
boundary state of this 1d SPT phase is given in Eq.~\ref{o3wf}.
Under U(1) and $Z_2$ transformation, the boundary doublet $U$
transforms as \beqn U(1): U \rightarrow e^{i\theta \sigma^z/2} U,
\ \ \ Z_2 : U \rightarrow \sigma^x U. \eeqn

In 3d, there is also only one way of assigning the symmetry to the
O(5) vector: \beqn U(1)&:& (n_1 + in_2) \rightarrow e^{i\theta}
(n_1 + in_2), \ \ n_b \rightarrow n_b, \ b = 3, 4, 5; \cr\cr Z_2
&:& n_1 \rightarrow n_1, \ \ n_b, \rightarrow - n_b, \ \ b = 2,
\cdots 5.\eeqn In both 1d and 3d, $\Theta = 4\pi$ is equivalent to
$\Theta = 0$, thus in both 1d and 3d the classification is
$\mathbb{Z}_2$.

In 2d, there are two independent ways of assigning $U(1) \rtimes
Z_2$ transformations to the O(4) vector $\vec{n}$: \beqn (1) &:&
U(1): (n_1 + in_2) \rightarrow e^{i\theta} (n_1 + in_2), \cr \cr
&& (n_3 + in_4) \rightarrow e^{i\theta} (n_3 + in_4) ; \cr\cr &&
Z_2 : n_1, n_3 \rightarrow n_1,n_3, \ \ n_2, n_4 \rightarrow -
n_2, - n_4; \cr\cr (2) &:& U(1): \vec{n} \rightarrow \vec{n}, \ \
Z_2 : \vec{n} \rightarrow - \vec{n}.\eeqn The transformation (1)
contributes $\mathbb{Z}$ classification, while transformation (2)
contributes $\mathbb{Z}_2$ classification, $i.e.$ in 2d the
classification is $\mathbb{Z} \times \mathbb{Z}_2$. {\it The final
classification of SPT phases with $U(1) \rtimes Z_2$ symmetry is}:
\beqn 1d: \mathbb{Z}_2, \ \ \ 2d: \mathbb{Z} \times \mathbb{Z}_2,
\ \ \ 3d: \mathbb{Z}_2. \eeqn

\subsection{SPT phases with $U(1) \times Z_2$ symmetry}

In both 1d and 3d, there is no way of assigning $U(1) \times Z_2$
transformations to vector $\vec{n}$ that keeps the $\Theta$ term
invariant. But in 2d, we can construct three root phases: \beqn
(1) &:& U(1): (n_1 + in_2) \rightarrow e^{i\theta} (n_1 + in_2),
\cr \cr && (n_3 + in_4) \rightarrow e^{i\theta} (n_3 + in_4) ;
\cr\cr && Z_2 : \vec{n} \rightarrow \vec{n}; \cr\cr (2) &:& U(1):
\vec{n} \rightarrow \vec{n}, \ \ Z_2 : \vec{n} \rightarrow -
\vec{n}; \cr\cr (3) &:& U(1): (n_1 + in_2) \rightarrow e^{i\theta}
(n_1 + in_2), \cr \cr && n_{3,4} \rightarrow n_{3,4}; \cr\cr &&
Z_2: n_{1,2} \rightarrow n_{1,2}, \ \ \ n_{3,4} \rightarrow -
n_{3,4}. \label{2du1z2} \eeqn The first transformation contributes
classification $\mathbb{Z}$, while transformations (2) and (3)
both contribute classification $\mathbb{Z}_2$, {\it thus the final
classification of SPT phases with $U(1) \times Z_2$ symmetry is}:
\beqn 1d: \mathbb{Z}_1, \ \ \ 2d: \mathbb{Z} \times
(\mathbb{Z}_2)^2, \ \ \ 3d: \mathbb{Z}_1. \eeqn

\subsection{SPT phases with $U(1) \rtimes Z_2^T$ symmetry}

A boson operator $b$ with $U(1) \rtimes Z_2^T$ symmetry transforms
as $b\rightarrow b$ under $Z_2^T$. In 1d, the only $U(1) \rtimes
Z_2^T$ symmetry transformation that keeps Eq.~\ref{o3theta}
invariant is the same transformation as $Z_2^T$ SPT phase, namely
vector $\vec{n}$ does not transform under $U(1)$, but changes sign
under $Z_2^T$.

In 2d, the only transformation that keeps Eq.~\ref{o4theta}
invariant is \beqn U(1)&:& (n_1 + in_2) \rightarrow e^{i\theta}
(n_1 + in_2), \ n_{3,4} \rightarrow n_{3,4}; \cr\cr  Z_2^T &:& n_1
\rightarrow n_1, \ \ n_a\rightarrow -n_a (a = 2,3,4), \eeqn and
this NLSM gives classification $\mathbb{Z}_2$.

The NLSMs for $U(1) \rtimes Z_2^T$ SPT phases in 3d have been
discussed in Ref.~\onlinecite{senthilashvin}, and in 3d the
classification is $(\mathbb{Z}_2)^2$. {\it Thus the final
classification of SPT phases with $U(1) \rtimes Z_2^T$ symmetry
is}: \beqn 1d: \mathbb{Z}_2, \ \ \ 2d: \mathbb{Z}_2, \ \ \ 3d:
(\mathbb{Z}_2)^2. \eeqn

\subsection{SPT phases with $U(1) \times Z_2^T$ symmetry}

In 1d, there are two independent transformations that keep
Eq.~\ref{o3theta} invariant: \beqn (1) &:& U(1): (n_1 + in_2)
\rightarrow e^{i\theta} (n_1 + in_2), \ \ n_3 \rightarrow n_3;
\cr\cr  && Z_2^T : n_{1,2} \rightarrow n_{1,2}, \ \ \ n_3
\rightarrow - n_3, \cr\cr (2) &:& U(1) : \vec{n} \rightarrow
\vec{n}, \cr\cr && Z_2^T: \vec{n} \rightarrow - \vec{n}. \eeqn In
2d there is no $U(1) \times Z_2^T$ transformation that keeps
Eq.~\ref{o4theta} invariant. In 3d the NLSMs for $U(1) \times
Z_2^T$ SPT phases were discussed in
Ref.~\onlinecite{senthilashvin}. {\it The final classification of
SPT phases with $U(1) \times Z_2^T$ symmetry is}: \beqn 1d:
(\mathbb{Z}_2)^2, \ \ \ 2d: \mathbb{Z}_1, \ \ \ 3d:
(\mathbb{Z}_2)^3. \eeqn

\subsection{SPT phases with $Z_2 \times Z_2$ symmetry}

In 1d, there is only one $Z_2 \times Z_2$ transformation that
keeps Eq.~\ref{o3theta} invariant: \beqn Z_2^A &:&
n_{1,2}\rightarrow - n_{1,2}, \ \ n_3 \rightarrow n_3, \cr\cr
Z_2^B &:& n_1 \rightarrow n_1, \ \ n_{2,3} \rightarrow -n_{2,3}.
\eeqn The boundary state $U$ defined in Eq.~\ref{o3wf} transforms
as \beqn Z_2^A: U \rightarrow i\sigma^z U, \ \ \ Z_2^B : U
\rightarrow \sigma^x U. \eeqn Thus $Z_2^A$ and $Z_2^B$ no longer
commute with each other at the boundary.

In 2d, there are three independent $Z_2 \times Z_2$
transformations (three different root phases): \beqn (1) &:& Z_2^A
: \vec{n} \rightarrow - \vec{n}, \ \ Z_2^B: \vec{n} \rightarrow
\vec{n}; \cr\cr (2) &:& Z_2^A : \vec{n} \rightarrow \vec{n}, \ \
Z_2^B: \vec{n} \rightarrow -\vec{n}; \cr\cr (3) &:& Z_2^A :
n_{1,2} \rightarrow - n_{1,2}, \ \ n_{3 ,4} \rightarrow n_{3,4};
\cr\cr && Z_2^B : n_{1, 2} \rightarrow n_{1,2}, \ \ n_{3,4}
\rightarrow -n_{3,4}. \eeqn

In 3d, there are also two independent $Z_2 \times Z_2$
transformations that keep Eq.~\ref{o5theta} invariant (two root
phases): \beqn (1) &:& Z_2^A : n_{1, 2} \rightarrow - n_{1, 2}, \
\ n_a \rightarrow n_a (a = 3,4,5); \cr\cr && Z_2^B :
n_1,\rightarrow n_1, \ \ n_a \rightarrow - n_a (a = 2, \cdots 5);
\cr\cr (2) &:& Z_2^B : n_{1, 2} \rightarrow - n_{1, 2}, \ \ n_a
\rightarrow n_a (a = 3,4,5); \cr\cr && Z_2^A : n_1,\rightarrow
n_1, \ \ n_a \rightarrow - n_a (a = 2, \cdots 5). \eeqn

As we discussed in section II.F, the boundary of these 3d SPT
phases can have 2d $Z_2$ topological order. A 2d $Z_2$ topological
phase has $e$ and $m$ anyon excitations, and these anyons
correspond to vortices of certain components of order parameter
$\vec{n}$. If the $e$ and $m$ anyons correspond to vortices of
$(n_3, n_4)$ and $(n_1, n_2)$ respectively, then according to
Eq.~\ref{0dwzw}, the $e$ excitation corresponds to a $0+1d$ O(3)
WZW model for vector $(n_1, n_2, n_5)$, and the $m$ excitation
corresponds to a $0+1d$ WZW model for vector $(n_3, n_4, n_5)$.
The boundary anyons of phase $(1)$ transform as: \beqn (1) &:&
Z_2^A: U_e \rightarrow i\sigma^z U_e, \ \ \ U_m \rightarrow U_m;
\cr\cr && Z_2^B: U_e \rightarrow \sigma^x U_e, \ \ \ U_m
\rightarrow i\sigma^y U_m^\ast. \label{z2z2anyon}\eeqn Notice that
under $Z_2^B$, a vortex of $(n_1, n_2)$ becomes an antivortex,
thus the transformation of $U_m$ under $Z_2^B$ involves a complex
conjugation. The transformation of boundary anyons of phase (2) is
the same as Eq.~\ref{z2z2anyon} after interchanging $Z_2^A$ and
$Z_2^B$.

{\it The final classification of SPT phases with $Z_2 \times Z_2$
symmetry is}: \beqn 1d: \mathbb{Z}_2, \ \ \ 2d: (\mathbb{Z}_2)^3,
\ \ \ 3d: (\mathbb{Z}_2)^2. \eeqn

\subsection{SPT phases with $Z_2 \times Z_2^T$ symmetry}

In 1d and 3d, the SPT phases with $Z_2 \times Z_2^T$ symmetry are
simply SPT phases with $U(1) \times Z_2^T$ symmetry after reducing
U(1) to its subgroup $Z_2$. The classification is the same as the
$U(1) \times Z_2^T$ SPT phases discussed in the previous
subsection. In 2d, there are two different root phases that
correspond to the following transformations: \beqn (1) &:& Z_2:
n_{1,2} \rightarrow -n_{1,2}, \ \ n_{3,4} \rightarrow n_{3,4},
\cr\cr && Z_2^T : n_1 \rightarrow n_1, \ \ n_a \rightarrow -n_a (a
= 2,3,4); \cr\cr (2) &:& Z_2 : \vec{n} \rightarrow - \vec{n},
\cr\cr && Z_2^T : n_1 \rightarrow n_1, \ \ n_a \rightarrow -n_a (a
= 2,3,4). \eeqn

{\it The final classification of SPT phases with $Z_2 \times
Z_2^T$ symmetry is}: \beqn 1d: (\mathbb{Z}_2)^2, \ \ \ 2d:
(\mathbb{Z}_2)^2, \ \ \ 3d: (\mathbb{Z}_2)^3. \eeqn

\subsection{SPT phases with $Z_m$ symmetry}

In 1d and 3d, there are no nontrivial $Z_m$ transformations that
can keep Eq.~\ref{o3theta} and Eq.~\ref{o5theta} invariant. In 2d,
we can construct the following root phase: \beqn Z_m &:& (n_1 +
in_2) \rightarrow e^{i 2\pi k/m} (n_1 + in_2); \cr\cr && (n_3 +
in_4) \rightarrow e^{i 2\pi k/m} (n_3 + in_4), \cr\cr && k = 0,
\cdots m-1 \eeqn Using the method in section II, we can
demonstrate that with these transformations, Eq.~\ref{o4theta}
with $\Theta = 2\pi m$ and $\Theta = 0$ are equivalent to each
other, thus the classification is $\mathbb{Z}_m$ in 2d.

{\it The final classification of SPT phases with $Z_m$ symmetry
is}: \beqn 1d: \mathbb{Z}_1, \ \ \ 2d: \mathbb{Z}_m, \ \ \ 3d:
\mathbb{Z}_1. \eeqn

\subsection{SPT phases with $Z_m \rtimes Z_2$ symmetry}

In 1d, there is one SPT phase with $U(1)\rtimes Z_2$ symmetry.
Naively one would expect that when U(1) is broken down to $Z_m$,
this SPT phase survives and becomes a SPT phase with $Z_m \rtimes
Z_2$ symmetry. However, this statement is only true for even $m$,
and when $m$ is odd the $U(1)\rtimes Z_2$ SPT phase becomes
trivial once U(1) is broken down to $Z_m$.

The 1d $U(1)\rtimes Z_2$ SPT phase is described by a 1d O(3) NLSM
of vector $\vec{n}$ with $\Theta = 2\pi$, and $B \sim (n_1 +
in_2)$ is a charge-1 boson under the U(1) rotation. Because the
classification of 1d $U(1)\rtimes Z_2$ SPT phase is
$\mathbb{Z}_2$, $\Theta = 2\pi$ is equivalent to $\Theta = 2\pi m$
for odd $m$. As we discussed in section IID, this NLSM with
$\Theta = 2\pi m$ is equivalent to another NLSM defined with
$\vec{n}^\prime$ and $\Theta = 2\pi$, where $B^\prime \sim
(n_1^\prime + i n_2^\prime) \sim (n_1 + in_2)^m$ is a charge-$m$
boson. Under $Z_2$ transformation, $n_1^\prime \rightarrow
n_1^\prime$, $n_2^\prime \rightarrow - n_2^\prime$.

Now let us break U(1) down to its subgroup $Z_m$. $B^\prime$
transforms trivially under $Z_m$, thus we are allowed to turn on a
Zeeman term $\mathrm{Re}[B^\prime] \sim n_1^\prime $ which fully
polarizes $n_1^\prime$ and kills the SPT phase. Thus the original
$U(1)\rtimes Z_2$ SPT phase is instable under U(1) to $Z_m$
breaking with odd $m$.

The discussion above is very abstract, let us understand this
result physically, and we will take $m = 3$ as an example. With a
full SO(3) symmetry and $\Theta = 2\pi$ in the bulk, the ground
state of the boundary is a spin-1/2 doublet in Eq.~\ref{o3wf}. The
excited states of the boundary include a spin-3/2 quartet. When
$\Theta = 6\pi$ in the bulk, the boundary ground state is a
spin-3/2 quartet. The spin-3/2 and spin-1/2 states can have a
boundary transition (level crossing at the boundary) without
closing the bulk gap, thus $\Theta = 2\pi$ and $6\pi$ are
equivalent in the bulk. Now let us take $\Theta = 6\pi$ in the
bulk, and break the SO(3) down to $Z_3 \rtimes Z_2$. Then we are
allowed to turn on a perturbation $\cos(3\phi)$ at the boundary
(which precisely corresponds to the Zeeman coupling
$\mathrm{Re}[B^\prime] \sim n_1^\prime $ discussed in the previous
paragraph), which will mix and split the two states $S^z = \pm
3/2$ at the boundary, and the boundary ground state can become
nondegenerate. Thus when $m$ is odd, the $U(1)\rtimes Z_2$ SPT
phase does not survive the symmetry breaking from U(1) to $Z_m$.

The same situation occurs in 2d and 3d. There is a 3d SPT phase
with $U(1) \rtimes Z_2$ symmetry, but once we break the U(1) down
to $Z_m$, this SPT phase does not survive when $m$ is odd. When
$m$ is even, besides the phase deduced from $U(1) \rtimes Z_2$ SPT
phase, one can construct another root phase: \beqn Z_2 &:& n_{1,2}
\rightarrow - n_{1,2}, \ \ n_a \rightarrow n_a \ (a = 3,4,5);
\cr\cr Z_m &:& n_1,\rightarrow n_1, \ \ n_a \rightarrow (-1)^k n_a
\ (a = 2, \cdots 5), \cr\cr && k = 0, \cdots m-1. \label{3dzmrz2}
\eeqn Here $n_a (a = 2, \cdots 5)$ still carries a nontrivial
representation of $Z_m$ for even integer $m$.
$n_a$ with $a = 3,4,5$ can be viewed as the real parts of
charge-$m/2$ bosons, while $n_2$ is the imaginary part of such
charge-$m/2$ boson. This construction does not apply for odd $m$.

In 2d, for arbitrary $m > 1$, the $U(1) \rtimes Z_2$ SPT phases
survive under $U(1)$ to $Z_m$ symmetry breaking. 
With even $m$, another root phase can be constructed \beqn Z_m &:&
n_{1,2} \rightarrow (-1)^k n_{1,2}, \ \ n_{3,4} \rightarrow
n_{3,4}; \cr\cr Z_2 &:& n_{1,2} \rightarrow n_{1,2}, \ \ n_{3,4}
\rightarrow - n_{3,4}, \cr\cr && k = 0, \cdots m-1. \eeqn Here
$n_1$ and $n_2$ are both the real parts of the charge-$m/2$
bosons.

{\it The final classification of SPT phases with $Z_m \rtimes Z_2$
symmetry is}: \beqn 1d: \mathbb{Z}_{(2,m)}, \ \ \ 2d: \mathbb{Z}_m
\times \mathbb{Z}_2 \times \mathbb{Z}_{(2,m)}, \ \ \ 3d:
(\mathbb{Z}_{(2,m)})^2. \eeqn

\subsection{SPT phases with $Z_m \times Z_2$ symmetry}

The case $m = 2$ has already been discussed. When $m > 2$, one
would naively expect these SPT phases can be interpreted as $U(1)
\times Z_2$ SPT phases after breaking U(1) to its $Z_m$ subgroup,
but again this is not entirely correct. In 1d there is no SPT
phase with $U(1)\times Z_2$ symmetry, simply because we cannot
find a nontrivial transformation of $\vec{n}$ under $U(1)\times
Z_2$ that keeps Eq.~\ref{o3theta} invariant. But when $m$ is an
even number, we can construct one SPT phase with $Z_m \times Z_2$
symmetry using Eq.~\ref{o3theta}: \beqn Z_m &:& n_{1,2}
\rightarrow (-1)^k n_{1,2}, \ \ n_3 \rightarrow n_3, \cr\cr Z_2
&:& n_1 \rightarrow n_1, \ \ n_{2,3} \rightarrow - n_{2,3}, \cr\cr
&& k = 0, \cdots m-1. \eeqn The $Z_m$ and $Z_2$ transformations on
$\vec{n}$ commute with each other.

Again this construction applies to even integer $m$ only. The
boundary states of this 1d SPT phase have the following
transformations: \beqn Z_m &:& U \rightarrow (i \sigma^z)^k U, \ \
\ Z_2 : U \rightarrow \sigma^x U ; \cr\cr && k = 0, \cdots m-1.
\eeqn Thus the boundary states carry projective representations of
$Z_m\times Z_2$, and the transformations of $Z_m$ and $Z_2$ do not
commute.

Similar situations occur in 3d. In 3d, we can construct two root
phases for even $m$, even though there is no SPT phase with $U(1)
\times Z_2$ symmetry in 3d : \beqn (1) &:& Z_m : n_{1, 2}
\rightarrow (-1)^k n_{1,2}, \ \ n_a \rightarrow n_a (a = 3,4,5);
\cr\cr && Z_2 : n_1,\rightarrow n_1, \ \ n_a \rightarrow - n_a \
(a = 2, \cdots 5); \cr\cr (2) &:& Z_2 : n_{1, 2} \rightarrow -
n_{1,2}, \ \ n_a \rightarrow n_a (a = 3,4,5); \cr\cr && Z_m :
n_1,\rightarrow n_1, \ \ n_a \rightarrow (-1)^k n_a (a = 2, \cdots
5) ; \cr\cr && k = 0, \cdots m-1. \label{3dzmz2}\eeqn The boundary
of these 3d SPT phases can have 2d $Z_2$ topological order. If the
$e$ and $m$ anyons correspond to vortices of $(n_3, n_4)$ and
$(n_1, n_2)$ respectively, then the boundary anyons of phase $(1)$
transform as: \beqn (1) &:& Z_m: U_e \rightarrow (i\sigma^z)^k
U_e, \ \ \ U_m \rightarrow U_m; \cr\cr && Z_2: U_e \rightarrow
\sigma^x U_e, \ \ \ U_m \rightarrow i\sigma^y U_m^\ast. \eeqn The
transformation of boundary anyons of phase $(2)$ can be derived in
the same way.

In 2d all the $Z_m \times Z_2$ SPT phases can be deduced from
$U(1) \times Z_2$ SPT phases, by breaking U(1) down to its $Z_m$
subgroup. Thus cases (1), (2) and (3) in Eq.~\ref{2du1z2} seem to
reduce to SPT phases with $Z_m \times Z_2$ symmetry after breaking
U(1) down to $Z_m$. However, case (3) in Eq.~\ref{2du1z2} becomes
the trivial phase when $m$ is odd. In case (3) of $U(1) \times
Z_2$ SPT phase (Eq.~\ref{2du1z2}), the NLSM is constructed with a
charge-1 boson $B \sim (n_1 + in_2)$, and because case (3)
contributes classification $\mathbb{Z}_2$, $\Theta = 2\pi m$ is
equivalent to $\Theta  = 2\pi$ for odd $m$. Also, the NLSM with
$\Theta = 2\pi m$ is equivalent to the NLSM with $\Theta = 2\pi$
constructed using a charge-$m$ boson $B^\prime \sim (n_1^\prime +
i n_2^\prime) \sim (n_1 + in_2)^m$. Now let us break the U(1)
symmetry down to $Z_m$. Because $B^\prime$ is invariant under
$Z_m$ and $Z_2$, we can turn on a linear Zeeman term that
polarizes $\mathrm{Re}[B^\prime] \sim n_1^\prime$, and destroy the
boundary states. Thus the NLSM constructed with the charge-$m$
boson $B^\prime$ is trivial once we break U(1) down to $Z_m$. This
implies that when $m$ is odd, case (3) in Eq.~\ref{2du1z2} becomes
a trivial phase once U(1) is broken down to $Z_m$.

{\it The final classification of SPT phases with $Z_m \times Z_2$
symmetry is}: \beqn 1d: \mathbb{Z}_{(2,m)}, \ \ \ 2d: \mathbb{Z}_m
\times \mathbb{Z}_2 \times \mathbb{Z}_{(2,m)}, \ \ \ 3d:
(\mathbb{Z}_{(2,m)})^2. \eeqn

\subsection{SPT phases with $Z_m \rtimes Z_2^T$ symmetry}

Again, the situation depends on the parity of $m$. If $m$ is odd,
then in 1d and 3d the only SPT phase is the SPT phase with $Z_2^T$
only. In 2d and 3d the $U(1) \rtimes Z_2^T$ SPT phases (except for
the one with $Z_2^T$ symmetry only) do not survive when U(1) is
broken down to $Z_m$ with odd $m$. The reason is similar to what
we discussed in the previous two subsections.

When $m$ is even, then in 1d besides the Haldane phase with
$Z_2^T$ symmetry, we can construct another SPT phase: \beqn Z_m
&:& n_{1, 2} \rightarrow (-1)^k n_{1,2}, \ \ n_3 \rightarrow n_3,
\cr \cr && k = 0, \cdots m-1;  \cr\cr Z_2^T &:& \vec{n}
\rightarrow -\vec{n}. \eeqn Here $n_1$ and $n_2$ are both
imaginary parts of charge-$m/2$ bosons. The boundary state is a
Kramers doublet and transforms as \beqn Z_m &:& U \rightarrow
(i\sigma^z)^k U, \ \ \ Z_2^T : U \rightarrow i\sigma^y U ; \cr\cr
&& k = 0, \cdots m-1. \label{1dzmrz2t}\eeqn


In 2d, we can construct two different root phases: \beqn (1) &&
Z_m : (n_1 + i n_2) \rightarrow (n_1 + in_2)e^{i2\pi k /m}, \cr
\cr && n_3, n_4 \rightarrow n_3, n_4; \cr \cr && Z_2^T : n_1
\rightarrow n_1, \ \ n_a \rightarrow -n_a (a = 2,3,4); \cr\cr (2)
&& Z_m : \vec{n} \rightarrow (-1)^k \vec{n}; \cr \cr && Z_2^T :
n_1 \rightarrow n_1, \ \ n_a \rightarrow -n_a (a = 2,3,4); \cr\cr
&& k = 0, \cdots m-1. \eeqn Phase (1) is the same phase as the 2d
$U(1) \rtimes Z_2^T$ SPT phase, after breaking U(1) to $Z_m$;
phase (2) is a new phase, where $n_1$ is the real part of a
charge-$m/2$ boson, while $n_{2,3,4}$ are the imaginary parts of
such charge-$m/2$ bosons.

Using similar methods, we can construct three root phases in 3d
for even $m$. Two of the phases can be deduced from the 3d $U(1)
\rtimes Z_2^T$ SPT phases. The third root phase has the following
transformation: \beqn Z_m &:& n_{1,2} \rightarrow (-1)^k n_{1, 2},
\ \ n_a \rightarrow n_a (a = 3,4,5); \cr\cr Z_2^T &:& \vec{n}
\rightarrow - \vec{n} ; \cr\cr && k = 0, \cdots m-1.
\label{3dzmrz2t}\eeqn Both $n_1$ and $n_2$ are imaginary parts of
charge-$m/2$ bosons.

Just like the 3d SPT phase with $U(1) \rtimes Z_2^T$ symmetry, the
2d boundary of the 3d $Z_m \rtimes Z_2^T$ SPT phase described by
Eq.~\ref{3dzmrz2t} can have a $Z_2$ topological order with
electric and magnetic anyons. The electric and magnetic anyons are
both Kramers doublet, and only one of them has a nontrivial
transformation under $Z_m$: $Z_m : U \rightarrow (i\sigma^z)^k U$,
$(k = 0, \cdots m-1)$.


{\it The final classification of SPT phases with $Z_m \rtimes
Z^T_2$ symmetry is}: \beqn 1d: \mathbb{Z}_2 \times
\mathbb{Z}_{(2,m)}, \ \ \ 2d: (\mathbb{Z}_{(2,m)})^2, \ \ \ 3d:
\mathbb{Z}_2 \times (\mathbb{Z}_{(2,m)})^2. \eeqn

\subsection{SPT phases with $Z_m \times Z_2^T$ symmetry}

In 1d and 3d, the SPT phases with $Z_m \times Z_2^T$ symmetry can
all be deduced from $U(1) \times Z_2^T$ symmetry by breaking U(1)
down to $Z_m$. Again, when $m$ is odd, some of the SPT phases
become trivial, for the same reason as what we discussed before.

In 2d there is no SPT phase with $U(1) \times Z_2^T$ symmetry, but
when $m$ is even we can construct two root phases, which {\it
cannot} be deduced from $U(1) \times Z_2^T$ SPT phases: \beqn (1)
&:& Z_m : \vec{n} \rightarrow (-1)^k \vec{n}; \cr \cr  && Z_2^T :
n_1 \rightarrow n_1, \ \ \ n_a \rightarrow - n_a (a = 2,3,4);
\cr\cr (2) &:& Z_m : n_{1,2} \rightarrow (-1)^k n_{1,2}, \ \
n_{3,4} \rightarrow n_{3,4}; \cr \cr && Z_2^T : n_1 \rightarrow
n_1, \ \ \ n_a \rightarrow - n_a (a = 2,3,4) ; \cr\cr && k = 0,
\cdots m-1. \eeqn

{\it The final classification of SPT phases with $Z_m \times
Z^T_2$ symmetry is}: \beqn 1d: \mathbb{Z}_2 \times
\mathbb{Z}_{(2,m)}, \ \ \ 2d: (\mathbb{Z}_{(2,m)})^2, \ \ \ 3d:
\mathbb{Z}_2 \times (\mathbb{Z}_{(2,m)})^2. \eeqn

\subsection{SPT phases with $SO(3)$ symmetry}

In 1d, the SO(3) symmetry leads to the Haldane phase, which is
described by Eq.~\ref{o3theta} with $\Theta = 2\pi$. In 3d, there
is no way to assign SO(3) symmetry to the five-component vector
$\vec{n}$ which makes the $\Theta-$term invariant, thus there is
no 3d SPT phase with SO(3) symmetry.

In 2d, Ref.~\onlinecite{liuwen} has given a nice way of describing
SPT phase with SO(3) symmetry, which is a principal chiral model
defined with group elements $SO(3)$. We will argue that the SO(3)
principal chiral model in Ref.~\onlinecite{liuwen} can be formally
rewritten as the O(4) NLSM Eq.~\ref{o4theta}, because we can
represent every group element $G_{ab}$ ($ 3 \times 3$ orthogonal
matrix) as a SU(2) matrix $\mathcal{Z}$: \beqn G_{ab} =
\frac{1}{2} \mathrm{tr}[\mathcal{Z}^\dagger \sigma^a \mathcal{Z}
\sigma^b], \label{z}\eeqn and the SU(2) matrix $\mathcal{Z}$ is
equivalent to an O(4) vector $\vec{n}$ with unit length:
$\mathcal{Z} = n^4 I_{2\times 2} + i \vec{n} \cdot \vec{\sigma}$.
We propose that the minimal SO(3) SPT phase discussed in
Ref.~\onlinecite{liuwen} can be effectively described by
Eq.~\ref{o4theta} with $\Theta = 8\pi$: \beqn \mathcal{S}_{2d} &=&
\int d^2x d\tau \ \frac{1}{g} (\partial_\mu \vec{n})^2 + \frac{i 8
\pi }{12 \pi^2} \epsilon_{abcd}\epsilon_{\mu\nu\rho} n^a
\partial_\mu n^b
\partial_\nu n^c \partial_\rho n^d \cr\cr &=&
\int d^2x d\tau \ \frac{1}{g} \mathrm{tr} [\partial_\mu
\mathcal{Z}^\dagger
\partial_\mu \mathcal{Z}] + \frac{i8\pi}{24\pi^2}
\mathrm{tr}[(\mathcal{Z}^\dagger d \mathcal{Z})^3].
\label{o4theta2} \eeqn Physically, Eq.~\ref{o4theta2} with $\Theta
= 8\pi$ gives SU(2) Hall conductivity $\sigma_{SU(2)} = 8$, or
equivalently SO(3) Hall conductivity $\sigma_{SO(3)} = 2$, which
is the same as the principal chiral model in
Ref.~\onlinecite{liuwen}. Mathematically, when field $\mathcal{Z}$
has a instanton number $\int d^3x \
\mathrm{tr}[(\mathcal{Z}^\dagger d \mathcal{Z})^3]/(24\pi^2) = +1
$ in the 2+1d space-time, the SO(3) matrix field $G_{ab}$ defined
in Eq.~\ref{z} will have instanton number $\int d^3x \
\mathrm{tr}[(G^{-1} d G)^3]/(24\pi^2) = +4 $. This factor of $4$
is precisely why $\Theta = 8\pi$ in Eq.~\ref{o4theta2}.

In order to represent $G_{ab}$ as $\mathcal{Z}$, we need to
introduce a $Z_2$ gauge field that couples to $\mathcal{Z}$,
because $\mathcal{Z}$ is a ``fractional" representation of
$G_{ab}$, and $G_{ab}$ is invariant under gauge transformation
$\mathcal{Z} \rightarrow - \mathcal{Z}$. In the language of
lattice gauge theory, our statement in the previous paragraph
implies that one of the possible confined phases of this $Z_2$
gauge field is trivial in the bulk without any extra symmetry
breaking or topological degeneracy, namely the vison (a dynamical
$Z_2$ $\pi-$flux coupled to $\mathcal{Z}$) in the bulk can
condensed without breaking any symmetry. This is indeed possible,
because if we weakly break the SU(2) symmetry down to U(1),
Eq.~\ref{o4theta2} describes a bosonic integer quantum Hall state
with Hall conductivity $8$. A $\pi-$flux in this system carries
charge $4$, which can be fully screened by four bosons, while
maintaining its bosonic statistics. Thus a vison can safely
condense in the bulk, confine the field $\mathcal{Z}$, and drive
the system into a SO(3) SPT phase.

{\it The final classification of SPT phases with $SO(3)$ symmetry
is}: \beqn 1d: \mathbb{Z}_2 , \ \ \ 2d: \mathbb{Z}, \ \ \ 3d:
\mathbb{Z}_1. \eeqn

\subsection{SPT phases with $SO(3) \times Z_2^T$ symmetry}

In 1d, there are two different SPT root phases with $SO(3) \times
Z_2^T$ symmetry, which correspond to the following transformations
of O(3) vector $\vec{n}$: \beqn (1)&:& SO(3): n_a \rightarrow
G_{ab}n_b, \ \ \ Z_2^T : \vec{n} \rightarrow - \vec{n};  \cr\cr
(2)&:& SO(3): \vec{n} \rightarrow \vec{n}, \ \ \ Z_2^T : \vec{n}
\rightarrow - \vec{n}. \eeqn

In 2d, the SPT phases with $SO(3) \times Z_2^T$ symmetry were
discussed in Ref.~\onlinecite{xu2dspt}, and it is described by
Eq.~\ref{o4theta} with transformation \beqn SO(3) &:& n_a
\rightarrow G_{ab}n_b (a, b = 1, 2, 3), \ \ n_4 \rightarrow n_4;
\cr \cr Z_2^T &:& n_a \rightarrow n_a (a = 1, 2, 3), \ \ \ n_4
\rightarrow - n_4. \eeqn

In 3d, there are three root phases for $SO(3) \times Z_2^T$ SPT
phases, two of which have the following field theory: \beqn (1)
&:& SO(3) : \vec{n} \rightarrow \vec{n}, \ \ \ Z_2^T : \vec{n}
\rightarrow - \vec{n}; \cr\cr (2) &:& SO(3) : n_a \rightarrow
G_{ab}n_b (a, b = 1, 2, 3), \ \ n_{4,5} \rightarrow n_{4,5} \cr
\cr && Z_2^T : \vec{n} \rightarrow -\vec{n}; \eeqn phase (1) is
simply the SPT phase with $Z_2^T$ symmetry only. After we break
the SO(3) symmetry down to its inplane O(2) subgroup, phase (2)
will reduce to a SPT phase with $U(1) \times Z_2^T$ symmetry
discussed in Ref.~\onlinecite{senthilashvin}, which is a phase
whose bulk vortex line is a 1d Haldane phase with $Z_2^T$
symmetry.

Besides the two phases discussed above, there should be another
root phase (3) that will reduce to the $U(1) \times Z_2^T$ SPT
phase whose boundary is a bosonic quantum Hall state with Hall
conductivity $\pm 1$, when time-reversal symmetry is broken at the
boundary~\cite{senthilashvin}. In the next two paragraphs we will
argue {\it without} proof that this third root phase can be
described by Eq.~\ref{o5theta} with the following definition and
transformation of O(5) vector order parameter $\vec{n}$: \beqn (3)
&:& \mathcal{Z} = n^4 I_{2\times 2} + \sum_{a = 1}^3 i n_a
\sigma^a, \cr \cr && Z_2^T : \mathcal{Z} \rightarrow i\sigma^y
\mathcal{Z}, \ \ \ n_5 \rightarrow -n_5; \cr\cr && \Theta = 8\pi \
\mathrm{in} \ \mathrm{bulk}. \label{3dso3z2t3}\eeqn Here
$\mathcal{Z}$ is still the ``fractional" representation of SO(3)
matrix $G_{ab}$ introduced in Eq.~\ref{z}. If we break the $Z_2^T$
symmetry at the boundary of phase (3), the boundary becomes a 2d
SO(3) SPT phase with SO(3) Hall conductivity $\pm$1 (when SO(3) is
broken to U(1), the boundary becomes a bosonic integer quantum
Hall state with Hall conductivity $\pm$1), thus it cannot be
realized in a pure 2d bosonic system without degeneracy.

In principle $\mathcal{Z}$ is still coupled to a $Z_2$ gauge
field. We propose that the confined phase of this $Z_2$ gauge
field is the desired $SO(3) \times Z_2^T$ SPT phase. In the
confined phase of a 3d $Z_2$ gauge field, the vison loops of the
$Z_2$ gauge field proliferate. Since the $Z_2$ gauge field is
coupled to the fractional field $\mathcal{Z}$, a vison loop of
this $Z_2$ gauge field is bound with a vortex loop of SO(3) matrix
field $G_{ab}$~\cite{xusachdev}, which is defined based on
homotopy group $\pi_1[SO(3)] = \mathbb{Z}_2$, thus the confined
phase of the $Z_2$ gauge field is a phase where the SO(3) vortex
loops proliferate. If we reduce the SO(3) symmetry down to its
inplane U(1) symmetry, the vison loop reduces to the vortex loop
of the U(1) phase. When a bulk vortex (vison) loop ends at the
boundary, it becomes a 2d vortex (vison). This 2d vortex is a
fermion, because according to the previous paragraph, once the
$Z_2^T$ is broken at the boundary, the boundary becomes a boson
quantum Hall state with Hall conductivity $\pm$1. This is
consistent with the results for $U(1)\times Z_2^T$ SPT phase
discussed in Ref.~\onlinecite{senthilashvin,xusenthil,maxfisher}.
Thus the SPT phase described by Eq.~\ref{3dso3z2t3} is a phase
where SO(3) vortex loops proliferate, and the SO(3) vortices at
the boundary are fermions.

{\it The final classification of SPT phases with $SO(3) \times
Z_2^T$ symmetry is}: \beqn 1d: (\mathbb{Z}_2)^2 , \ \ \ 2d:
\mathbb{Z}_2, \ \ \ 3d: (\mathbb{Z}_2)^3. \eeqn

\subsection{SPT phases with $Z_2 \times Z_2 \times Z_2$ symmetry}

In 1d, we can construct three different root phases: \beqn (1) &:&
Z^A_2 : n_{1,2} \rightarrow - n_{1,2}, \ \ n_3 \rightarrow n_3;
\cr\cr && Z^B_2 : n_{1} \rightarrow n_{1}, \ \ n_{2,3} \rightarrow
- n_{2,3}; \cr\cr && Z_2^C : \vec{n} \rightarrow \vec{n}; \cr\cr
(2) &:& Z^B_2 : n_{1,2} \rightarrow - n_{1,2}, \ \ n_3 \rightarrow
n_3; \cr\cr && Z^C_2 : n_{1} \rightarrow n_{1}, \ \ n_{2,3}
\rightarrow - n_{2,3}; \cr\cr && Z_2^A : \vec{n} \rightarrow
\vec{n}; \cr\cr (3) &:& Z^C_2 : n_{1,2} \rightarrow - n_{1,2}, \ \
n_3 \rightarrow n_3; \cr\cr && Z^A_2 : n_{1} \rightarrow n_{1}, \
\ n_{2,3} \rightarrow - n_{2,3}; \cr\cr && Z_2^B : \vec{n}
\rightarrow \vec{n}. \eeqn

In 2d there are seven different root phases: \beqn (1) &:& Z^A_2 :
\vec{n} \rightarrow - \vec{n}, \ \ Z_2^B, Z_2^C:
\vec{n}\rightarrow \vec{n}; \cr\cr (2) &:& Z^B_2 : \vec{n}
\rightarrow - \vec{n}, \ \ Z_2^C, Z_2^A: \vec{n}\rightarrow
\vec{n}; \cr\cr (3) &:& Z^C_2 : \vec{n} \rightarrow - \vec{n}, \ \
Z_2^A, Z_2^B: \vec{n}\rightarrow \vec{n}; \cr\cr (4) &:& Z^A_2 :
n_{1,2} \rightarrow - n_{1,2}, \ \ n_{3,4} \rightarrow n_{3,4};
\cr\cr && Z^B_2 : n_{1,2} \rightarrow n_{1,2}, \ \ n_{3,4}
\rightarrow - n_{3,4}; \cr\cr && Z_2^C : \vec{n} \rightarrow
\vec{n}; \cr\cr (5) &:& Z^A_2 : n_{1,2} \rightarrow - n_{1,2}, \ \
n_{3,4} \rightarrow n_{3,4}; \cr\cr && Z^C_2 : n_{1,2} \rightarrow
n_{1,2}, \ \ n_{3,4} \rightarrow - n_{3,4}; \cr\cr && Z_2^B :
\vec{n} \rightarrow \vec{n}; \cr\cr (6) &:& Z^A_2 : n_{1,2}
\rightarrow - n_{1,2}, \ \ n_{3,4} \rightarrow n_{3,4}; \cr\cr &&
Z^B_2 : n_{1,3} \rightarrow - n_{1,3}, \ \ n_{2,4} \rightarrow
n_{2,4}; \cr\cr && Z_2^C : n_{1,4} \rightarrow - n_{1,4}, \ \
n_{2,3} \rightarrow n_{2,3}; \cr\cr (7) &:& Z_2^A: n_{2,3}
\rightarrow - n_{2,3}, \ \ n_{1,4} \rightarrow n_{1,4} \cr\cr &&
Z_2^B: n_{1,2} \rightarrow - n_{1,2}, \ \ n_{3,4} \rightarrow
n_{3,4}, \cr\cr && Z_2^C: n_{1,2} \rightarrow n_{1,2}, \ \ n_{3,4}
\rightarrow - n_{3,4}. \eeqn

In 3d there are six different root phases: \beqn (1) &:& Z_2^A:
n_{1,2} \rightarrow - n_{1,2}, \ \ n_a \rightarrow n_a, (a =
3,4,5); \cr\cr && Z_2^B: n_{1} \rightarrow n_{1}, \ \ n_a
\rightarrow - n_a, (a = 2, \cdots 5); \cr\cr && Z_2^C : \vec{n}
\rightarrow \vec{n};  \cr\cr (2) &:& Z_2^B: n_{1,2} \rightarrow -
n_{1,2}, \ \ n_a \rightarrow n_a, (a = 3,4,5); \cr\cr && Z_2^A:
n_{1} \rightarrow n_{1}, \ \ n_a \rightarrow - n_a, (a = 2, \cdots
5); \cr\cr && Z_2^C : \vec{n} \rightarrow \vec{n}; \cr\cr (3) &:&
Z_2^B: n_{1,2} \rightarrow - n_{1,2}, \ \ n_a \rightarrow n_a, (a
= 3,4,5); \cr\cr && Z_2^C: n_{1} \rightarrow n_{1}, \ \ n_a
\rightarrow - n_a, (a = 2, \cdots 5); \cr\cr && Z_2^A : \vec{n}
\rightarrow \vec{n}; \cr\cr (4) &:& Z_2^C: n_{1,2} \rightarrow -
n_{1,2}, \ \ n_a \rightarrow n_a, (a = 3,4,5); \cr\cr && Z_2^B:
n_{1} \rightarrow n_{1}, \ \ n_a \rightarrow - n_a, (a = 2, \cdots
5); \cr\cr && Z_2^A : \vec{n} \rightarrow \vec{n}; \cr\cr (5) &:&
Z_2^A: n_{1,2} \rightarrow - n_{1,2}, \ \ n_a \rightarrow n_a, (a
= 3,4,5); \cr\cr && Z_2^C: n_{1} \rightarrow n_{1}, \ \ n_a
\rightarrow - n_a, (a = 2, \cdots 5); \cr\cr && Z_2^B : \vec{n}
\rightarrow \vec{n}; \cr\cr (6) &:& Z_2^C: n_{1,2} \rightarrow -
n_{1,2}, \ \ n_a \rightarrow n_a, (a = 3,4,5); \cr\cr && Z_2^A:
n_{1} \rightarrow n_{1}, \ \ n_a \rightarrow - n_a, (a = 2, \cdots
5); \cr\cr && Z_2^B : \vec{n} \rightarrow \vec{n}; \cr\cr (7) &:&
Z_2^A : n_{1,2} \ra - n_{1,2}, \ \ n_{3,4,5} \ra n_{3,4,5}; \cr\cr
&& Z_2^B : n_{2,3} \ra - n_{2,3}, \ \ n_{1,4,5} \ra n_{1,4,5};
\cr\cr && Z_2^C : n_{4,5} \ra - n_{4,5}, \ \ n_{1,2,3} \ra
n_{1,2,3}; \cr\cr (8) &:& Z_2^A : n_{1,2} \ra - n_{1,2}, \ \
n_{3,4,5} \ra n_{3,4,5}; \cr\cr && Z_2^C : n_{2,3} \ra - n_{2,3},
\ \ n_{1,4,5} \ra n_{1,4,5}; \cr\cr && Z_2^B : n_{4,5} \ra -
n_{4,5}, \ \ n_{1,2,3} \ra n_{1,2,3}. \label{3dz23}\eeqn

All the other SPT phases can be constructed with these root phases
above. Here we will show one construction explicitly. For example,
one may think the following state should also exist in $3d$: \beqn
\cr\cr Z_2^B: n_{1,2} \rightarrow - n_{1,2}, \ \ n_{3,4,5}
\rightarrow n_{3,4,5}, \cr\cr Z_2^C: n_{2,3} \rightarrow -
n_{2,3}, \ \ n_{1,4,5} \rightarrow n_{1,4,5}, \cr\cr Z_2^A:
n_{4,5} \rightarrow - n_{4,5}, \ \ n_{1,2,3} \rightarrow
n_{1,2,3}. \label{new}\eeqn But this state can be obtained by
``merging" state $(7)$ and $(8)$ in Eq.~\ref{3dz23}. First of all,
since $n^{(7)}_{1,3,5}$ transform exactly equivalently to
$n^{(8)}_{1,5,3}$ under all symmetries, we can turn on coupling
between $\vec{n}^{(7)}$ and $\vec{n}^{(8)}$ to make
$n^{(7)}_{1,3,5} = n^{(8)}_{1,5,3}$. Now without loss of
generality these two vectors can be written as \beqn \vec{n}^{(7)}
&=& (\cos\theta N_1, \ \sin \theta \cos \alpha^{(7)}, \ \cos\theta
N_2, \cr\cr && \sin \theta \sin \alpha^{(7)}, \ \cos\theta N_3);
\cr\cr \vec{n}^{(8)} &=& (\cos\theta N_1, \ \sin \theta \cos
\alpha^{(8)}, \ \cos\theta N_3, \cr\cr && \sin \theta \sin
\alpha^{(8)}, \ \cos\theta N_2); \eeqn where $\vec{N}$ is a unit
three-component vector. All the symmetries transformations act on
$\vec{N}$ and $\alpha^{(7)}$, $\alpha^{(8)}$, while $\theta$ is
invariant under all symmetries.

Now let us define a new vector $\vec{n}^{(9)}$ using the
parametrization of $\vec{n}^{(7)}$ and $\vec{n}^{(8)}$: \beqn
\vec{n}^{(9)} &=& (\cos\theta N_2, \ \sin\theta \cos(\alpha^{(7)}
+ \alpha^{(8)}), \ \cos \theta N_3, \cr\cr && \sin\theta
\sin(\alpha^{(7)} + \alpha^{(8)}), \ \cos\theta N_1); \eeqn
Obviously, the O(5) instanton number of $\vec{n}^{(9)}$ is exactly
the sum of instantons of $\vec{n}^{(7)}$ and $\vec{n}^{(8)}$. More
importantly, $\vec{n}^{(9)}$ transforms under all the symmetries
as Eq.~\ref{new}, and since it can be ``merged" from phase $(7)$
and $(8)$, it should not be viewed as an independent root phase.

{\it The final classification of SPT phases with $Z_2 \times Z_2
\times Z_2$ symmetry is}: \beqn 1d: (\mathbb{Z}_2)^3 , \ \ \ 2d:
(\mathbb{Z}_2)^7, \ \ \ 3d: (\mathbb{Z}_2)^8. \eeqn

\section{Summary and Comments}

In this work we systematically classified and described bosonic
SPT phases with a large set of physically relevant symmetries for
all physical dimensions. We have demonstrated that all the SPT
phases discussed in this paper can be described by three universal
NLSMs Eq.~\ref{o3theta}, \ref{o4theta} and \ref{o5theta}, and the
classification of these SPT phases based on NLSMs is completely
identical to the group cohomology
classification~\cite{wenspt,wenspt2}. However, we have not built
the general connection between these two classifications, and it
is likely that SPT phases with some other symmetry groups (for
example symmetry much larger than O($d+2$)) can no longer be
described by these three NLSMs any more. In
Ref.~\onlinecite{xu3dspt,xu2dspt}, SPT phases that involve a large
symmetry group PSU($N$)$ = SU(N)/Z_N$ were discussed, and in these
systems it was necessary to introduce NLSMs with a larger target
manifold. But it is likely that all the SPT phases with arbitrary
symmetry groups (continuous or discontinuous) can be described by
a NLSM with certain continuous target manifold.

As we already mentioned, now it is clear that there is a series of
BSPT states beyond the group cohomology classification, and a
generalized field theory description for such states will be given
in Ref.~\onlinecite{xubeyond}. Our NLSM can also be very
conveniently generalized to the cases that involve lattice
symmetry such as inversion, as was discussed in
Ref.~\onlinecite{youxuinversion}, as long as we require our order
parameter $\vec{n}$ transform nontrivially under lattice symmetry.
We leave a thorough study of SPT states involving lattice symmetry
to future studies.

Recently it was pointed out that after the $3d$ SPT state is
coupled to gauge field, the gauge defects, which in $3d$ can be
loop excitations, can have a novel loop braiding
statistics~\cite{levinloop}. In a recent work we showed that this
loop statistics can also be computed using our NLSM field theory
discussed in this work~\cite{xuloop}.

The authors are supported by the the David and Lucile Packard
Foundation and NSF Grant No. DMR-1151208. The authors especially
want to thank Meng Cheng, Xie Chen and Yuan-Ming Lu for very
helpful discussions.

\bibliography{class}

\begin{thebibliography}{61}
\expandafter\ifx\csname natexlab\endcsname\relax\def\natexlab#1{#1}\fi
\expandafter\ifx\csname bibnamefont\endcsname\relax
  \def\bibnamefont#1{#1}\fi
\expandafter\ifx\csname bibfnamefont\endcsname\relax
  \def\bibfnamefont#1{#1}\fi
\expandafter\ifx\csname citenamefont\endcsname\relax
  \def\citenamefont#1{#1}\fi
\expandafter\ifx\csname url\endcsname\relax
  \def\url#1{\texttt{#1}}\fi
\expandafter\ifx\csname urlprefix\endcsname\relax\def\urlprefix{URL }\fi
\providecommand{\bibinfo}[2]{#2}
\providecommand{\eprint}[2][]{\url{#2}}

\bibitem[{\citenamefont{Chen et~al.}(2013{\natexlab{a}})\citenamefont{Chen, Gu,
  Liu, and Wen}}]{wenspt}
\bibinfo{author}{\bibfnamefont{X.}~\bibnamefont{Chen}},
  \bibinfo{author}{\bibfnamefont{Z.-C.} \bibnamefont{Gu}},
  \bibinfo{author}{\bibfnamefont{Z.-X.} \bibnamefont{Liu}}, \bibnamefont{and}
  \bibinfo{author}{\bibfnamefont{X.-G.} \bibnamefont{Wen}},
  \bibinfo{journal}{Phys. Rev. B} \textbf{\bibinfo{volume}{87}},
  \bibinfo{pages}{155114} (\bibinfo{year}{2013}{\natexlab{a}}).

\bibitem[{\citenamefont{Chen et~al.}(2012)\citenamefont{Chen, Gu, Liu, and
  Wen}}]{wenspt2}
\bibinfo{author}{\bibfnamefont{X.}~\bibnamefont{Chen}},
  \bibinfo{author}{\bibfnamefont{Z.-C.} \bibnamefont{Gu}},
  \bibinfo{author}{\bibfnamefont{Z.-X.} \bibnamefont{Liu}}, \bibnamefont{and}
  \bibinfo{author}{\bibfnamefont{X.-G.} \bibnamefont{Wen}},
  \bibinfo{journal}{Science} \textbf{\bibinfo{volume}{338}},
  \bibinfo{pages}{1604} (\bibinfo{year}{2012}).

\bibitem[{\citenamefont{Kane and Mele}(2005{\natexlab{a}})}]{kane2005a}
\bibinfo{author}{\bibfnamefont{C.~L.} \bibnamefont{Kane}} \bibnamefont{and}
  \bibinfo{author}{\bibfnamefont{E.~J.} \bibnamefont{Mele}},
  \bibinfo{journal}{Physical Review Letter} \textbf{\bibinfo{volume}{95}},
  \bibinfo{pages}{226801} (\bibinfo{year}{2005}{\natexlab{a}}).

\bibitem[{\citenamefont{Kane and Mele}(2005{\natexlab{b}})}]{kane2005b}
\bibinfo{author}{\bibfnamefont{C.~L.} \bibnamefont{Kane}} \bibnamefont{and}
  \bibinfo{author}{\bibfnamefont{E.~J.} \bibnamefont{Mele}},
  \bibinfo{journal}{Physical Review Letter} \textbf{\bibinfo{volume}{95}},
  \bibinfo{pages}{146802} (\bibinfo{year}{2005}{\natexlab{b}}).

\bibitem[{\citenamefont{Bernevig et~al.}(2006)\citenamefont{Bernevig, Hughes,
  and Zhang}}]{bernevig2006}
\bibinfo{author}{\bibfnamefont{B.~A.} \bibnamefont{Bernevig}},
  \bibinfo{author}{\bibfnamefont{T.~L.} \bibnamefont{Hughes}},
  \bibnamefont{and} \bibinfo{author}{\bibfnamefont{S.-C.} \bibnamefont{Zhang}},
  \bibinfo{journal}{Science} \textbf{\bibinfo{volume}{314}},
  \bibinfo{pages}{1757} (\bibinfo{year}{2006}).

\bibitem[{\citenamefont{Fu et~al.}(2008)\citenamefont{Fu, Kane, and
  Mele}}]{fukane}
\bibinfo{author}{\bibfnamefont{L.}~\bibnamefont{Fu}},
  \bibinfo{author}{\bibfnamefont{C.~L.} \bibnamefont{Kane}}, \bibnamefont{and}
  \bibinfo{author}{\bibfnamefont{E.~J.} \bibnamefont{Mele}},
  \bibinfo{journal}{Phys. Rev. Lett.} \textbf{\bibinfo{volume}{98}},
  \bibinfo{pages}{106803} (\bibinfo{year}{2008}).

\bibitem[{\citenamefont{Moore and Balents}(2007)}]{moorebalents2007}
\bibinfo{author}{\bibfnamefont{J.~E.} \bibnamefont{Moore}} \bibnamefont{and}
  \bibinfo{author}{\bibfnamefont{L.}~\bibnamefont{Balents}},
  \bibinfo{journal}{Physical Review B} \textbf{\bibinfo{volume}{75}},
  \bibinfo{pages}{121306(R)} (\bibinfo{year}{2007}).

\bibitem[{\citenamefont{Roy}(2009)}]{roy2007}
\bibinfo{author}{\bibfnamefont{R.}~\bibnamefont{Roy}},
  \bibinfo{journal}{Physical Review B} \textbf{\bibinfo{volume}{79}},
  \bibinfo{pages}{195322} (\bibinfo{year}{2009}).

\bibitem[{\citenamefont{Xu and Moore}(2006)}]{xuedge}
\bibinfo{author}{\bibfnamefont{C.}~\bibnamefont{Xu}} \bibnamefont{and}
  \bibinfo{author}{\bibfnamefont{J.~E.} \bibnamefont{Moore}},
  \bibinfo{journal}{Phys. Rev. B} \textbf{\bibinfo{volume}{73}},
  \bibinfo{pages}{045322} (\bibinfo{year}{2006}).

\bibitem[{\citenamefont{Wu et~al.}(2006)\citenamefont{Wu, Bernevig, and
  Zhang}}]{wuedge}
\bibinfo{author}{\bibfnamefont{C.}~\bibnamefont{Wu}},
  \bibinfo{author}{\bibfnamefont{B.~A.} \bibnamefont{Bernevig}},
  \bibnamefont{and} \bibinfo{author}{\bibfnamefont{S.-C.} \bibnamefont{Zhang}},
  \bibinfo{journal}{Phys. Rev. Lett.} \textbf{\bibinfo{volume}{96}},
  \bibinfo{pages}{106401} (\bibinfo{year}{2006}).

\bibitem[{\citenamefont{Xu}(2010)}]{xu3dedge}
\bibinfo{author}{\bibfnamefont{C.}~\bibnamefont{Xu}}, \bibinfo{journal}{Phys.
  Rev. B} \textbf{\bibinfo{volume}{81}}, \bibinfo{pages}{020411}
  (\bibinfo{year}{2010}).

\bibitem[{\citenamefont{Wang et~al.}(2013{\natexlab{a}})\citenamefont{Wang,
  Potter, and Senthil}}]{wangsenthilTI}
\bibinfo{author}{\bibfnamefont{C.}~\bibnamefont{Wang}},
  \bibinfo{author}{\bibfnamefont{A.~C.} \bibnamefont{Potter}},
  \bibnamefont{and} \bibinfo{author}{\bibfnamefont{T.}~\bibnamefont{Senthil}},
  \bibinfo{journal}{arXiv:1306.3223}  (\bibinfo{year}{2013}{\natexlab{a}}).

\bibitem[{\citenamefont{Bonderson et~al.}(2013)\citenamefont{Bonderson, Nayak,
  and Qi}}]{qinayakTI}
\bibinfo{author}{\bibfnamefont{P.}~\bibnamefont{Bonderson}},
  \bibinfo{author}{\bibfnamefont{C.}~\bibnamefont{Nayak}}, \bibnamefont{and}
  \bibinfo{author}{\bibfnamefont{X.-L.} \bibnamefont{Qi}},
  \bibinfo{journal}{arXiv:1306.3230}  (\bibinfo{year}{2013}).

\bibitem[{\citenamefont{Metlitski
  et~al.}(2013{\natexlab{a}})\citenamefont{Metlitski, Kane, and
  Fisher}}]{fisherTI}
\bibinfo{author}{\bibfnamefont{M.~A.} \bibnamefont{Metlitski}},
  \bibinfo{author}{\bibfnamefont{C.~L.} \bibnamefont{Kane}}, \bibnamefont{and}
  \bibinfo{author}{\bibfnamefont{M.~P.~A.} \bibnamefont{Fisher}},
  \bibinfo{journal}{arXiv:1306.3286}  (\bibinfo{year}{2013}{\natexlab{a}}).

\bibitem[{\citenamefont{Chen et~al.}(2013{\natexlab{b}})\citenamefont{Chen,
  Fidkowski, and Vishwanath}}]{chenTI}
\bibinfo{author}{\bibfnamefont{X.}~\bibnamefont{Chen}},
  \bibinfo{author}{\bibfnamefont{L.}~\bibnamefont{Fidkowski}},
  \bibnamefont{and}
  \bibinfo{author}{\bibfnamefont{A.}~\bibnamefont{Vishwanath}},
  \bibinfo{journal}{arXiv:1306.3250}  (\bibinfo{year}{2013}{\natexlab{b}}).

\bibitem[{\citenamefont{Levin and Gu}(2012)}]{levingu}
\bibinfo{author}{\bibfnamefont{M.}~\bibnamefont{Levin}} \bibnamefont{and}
  \bibinfo{author}{\bibfnamefont{Z.-C.} \bibnamefont{Gu}},
  \bibinfo{journal}{Phys. Rev. B} \textbf{\bibinfo{volume}{86}},
  \bibinfo{pages}{115109} (\bibinfo{year}{2012}).

\bibitem[{\citenamefont{Levin and Senthil}(2013)}]{levinsenthil}
\bibinfo{author}{\bibfnamefont{M.}~\bibnamefont{Levin}} \bibnamefont{and}
  \bibinfo{author}{\bibfnamefont{T.}~\bibnamefont{Senthil}},
  \bibinfo{journal}{Phys. Rev. Lett.} \textbf{\bibinfo{volume}{110}},
  \bibinfo{pages}{046801} (\bibinfo{year}{2013}).

\bibitem[{\citenamefont{Levin and Stern}(2012)}]{levinstern}
\bibinfo{author}{\bibfnamefont{M.}~\bibnamefont{Levin}} \bibnamefont{and}
  \bibinfo{author}{\bibfnamefont{A.}~\bibnamefont{Stern}},
  \bibinfo{journal}{Phys. Rev. B} \textbf{\bibinfo{volume}{86}},
  \bibinfo{pages}{115131} (\bibinfo{year}{2012}).

\bibitem[{\citenamefont{Liu and Wen}(2013)}]{liuwen}
\bibinfo{author}{\bibfnamefont{Z.-X.} \bibnamefont{Liu}} \bibnamefont{and}
  \bibinfo{author}{\bibfnamefont{X.-G.} \bibnamefont{Wen}},
  \bibinfo{journal}{Phys. Rev. Lett.} \textbf{\bibinfo{volume}{110}},
  \bibinfo{pages}{067205} (\bibinfo{year}{2013}).

\bibitem[{\citenamefont{Lu and Vishwanath}(2012)}]{luashvin}
\bibinfo{author}{\bibfnamefont{Y.-M.} \bibnamefont{Lu}} \bibnamefont{and}
  \bibinfo{author}{\bibfnamefont{A.}~\bibnamefont{Vishwanath}},
  \bibinfo{journal}{Phys. Rev. B} \textbf{\bibinfo{volume}{86}},
  \bibinfo{pages}{125119} (\bibinfo{year}{2012}).

\bibitem[{\citenamefont{Vishwanath and Senthil}(2013)}]{senthilashvin}
\bibinfo{author}{\bibfnamefont{A.}~\bibnamefont{Vishwanath}} \bibnamefont{and}
  \bibinfo{author}{\bibfnamefont{T.}~\bibnamefont{Senthil}},
  \bibinfo{journal}{Phys. Rev. X} \textbf{\bibinfo{volume}{3}},
  \bibinfo{pages}{011016} (\bibinfo{year}{2013}).

\bibitem[{\citenamefont{Xu}(2013{\natexlab{a}})}]{xu3dspt}
\bibinfo{author}{\bibfnamefont{C.}~\bibnamefont{Xu}}, \bibinfo{journal}{Phys.
  Rev. B} \textbf{\bibinfo{volume}{87}}, \bibinfo{pages}{174412}
  (\bibinfo{year}{2013}{\natexlab{a}}).

\bibitem[{\citenamefont{Oon et~al.}(2012)\citenamefont{Oon, Cho, and
  Xu}}]{xu2dspt}
\bibinfo{author}{\bibfnamefont{J.}~\bibnamefont{Oon}},
  \bibinfo{author}{\bibfnamefont{G.~Y.} \bibnamefont{Cho}}, \bibnamefont{and}
  \bibinfo{author}{\bibfnamefont{C.}~\bibnamefont{Xu}},
  \bibinfo{journal}{arXiv:1212.1726}  (\bibinfo{year}{2012}).

\bibitem[{\citenamefont{Xu and Senthil}(2013)}]{xusenthil}
\bibinfo{author}{\bibfnamefont{C.}~\bibnamefont{Xu}} \bibnamefont{and}
  \bibinfo{author}{\bibfnamefont{T.}~\bibnamefont{Senthil}},
  \bibinfo{journal}{Phys. Rev. B} \textbf{\bibinfo{volume}{87}},
  \bibinfo{pages}{174412} (\bibinfo{year}{2013}).

\bibitem[{\citenamefont{Wang and Senthil}(2013)}]{wangsenthil}
\bibinfo{author}{\bibfnamefont{C.}~\bibnamefont{Wang}} \bibnamefont{and}
  \bibinfo{author}{\bibfnamefont{T.}~\bibnamefont{Senthil}},
  \bibinfo{journal}{Phys. Rev. B} \textbf{\bibinfo{volume}{87}},
  \bibinfo{pages}{235122} (\bibinfo{year}{2013}).

\bibitem[{\citenamefont{Wang et~al.}(2013{\natexlab{b}})\citenamefont{Wang,
  Potter, and Senthil}}]{wangpottersenthil}
\bibinfo{author}{\bibfnamefont{C.}~\bibnamefont{Wang}},
  \bibinfo{author}{\bibfnamefont{A.~C.} \bibnamefont{Potter}},
  \bibnamefont{and} \bibinfo{author}{\bibfnamefont{T.}~\bibnamefont{Senthil}},
  \bibinfo{journal}{arXiv:1306.3238}  (\bibinfo{year}{2013}{\natexlab{b}}).

\bibitem[{\citenamefont{Chen et~al.}(2014{\natexlab{a}})\citenamefont{Chen, Lu,
  and Vishwanath}}]{chenluashvin}
\bibinfo{author}{\bibfnamefont{X.}~\bibnamefont{Chen}},
  \bibinfo{author}{\bibfnamefont{Y.-M.} \bibnamefont{Lu}}, \bibnamefont{and}
  \bibinfo{author}{\bibfnamefont{A.}~\bibnamefont{Vishwanath}},
  \bibinfo{journal}{Nature Communications} \textbf{\bibinfo{volume}{5}},
  \bibinfo{pages}{3507} (\bibinfo{year}{2014}{\natexlab{a}}).

\bibitem[{\citenamefont{Metlitski
  et~al.}(2013{\natexlab{b}})\citenamefont{Metlitski, Kane, and
  Fisher}}]{maxfisher}
\bibinfo{author}{\bibfnamefont{M.~A.} \bibnamefont{Metlitski}},
  \bibinfo{author}{\bibfnamefont{C.~L.} \bibnamefont{Kane}}, \bibnamefont{and}
  \bibinfo{author}{\bibfnamefont{M.~P.~A.} \bibnamefont{Fisher}},
  \bibinfo{journal}{arXiv:1302.6535}  (\bibinfo{year}{2013}{\natexlab{b}}).

\bibitem[{\citenamefont{Ye and Wen}(2013{\natexlab{a}})}]{yewen1}
\bibinfo{author}{\bibfnamefont{P.}~\bibnamefont{Ye}} \bibnamefont{and}
  \bibinfo{author}{\bibfnamefont{X.-G.} \bibnamefont{Wen}},
  \bibinfo{journal}{Phys. Rev. B} \textbf{\bibinfo{volume}{87}},
  \bibinfo{pages}{195128} (\bibinfo{year}{2013}{\natexlab{a}}).

\bibitem[{\citenamefont{Ye and Wen}(2013{\natexlab{b}})}]{yewen2}
\bibinfo{author}{\bibfnamefont{P.}~\bibnamefont{Ye}} \bibnamefont{and}
  \bibinfo{author}{\bibfnamefont{X.-G.} \bibnamefont{Wen}},
  \bibinfo{journal}{arXiv:1303.3572}  (\bibinfo{year}{2013}{\natexlab{b}}).

\bibitem[{\citenamefont{Cheng and Gu}(2013)}]{chenggu}
\bibinfo{author}{\bibfnamefont{M.}~\bibnamefont{Cheng}} \bibnamefont{and}
  \bibinfo{author}{\bibfnamefont{Z.-C.} \bibnamefont{Gu}},
  \bibinfo{journal}{arXiv:1302.4803}  (\bibinfo{year}{2013}).

\bibitem[{\citenamefont{Haldane}(1983{\natexlab{a}})}]{haldane1}
\bibinfo{author}{\bibfnamefont{F.~D.~M.} \bibnamefont{Haldane}},
  \bibinfo{journal}{Phys. Lett. A} \textbf{\bibinfo{volume}{93}},
  \bibinfo{pages}{464} (\bibinfo{year}{1983}{\natexlab{a}}).

\bibitem[{\citenamefont{Haldane}(1983{\natexlab{b}})}]{haldane2}
\bibinfo{author}{\bibfnamefont{F.~D.~M.} \bibnamefont{Haldane}},
  \bibinfo{journal}{Phys. Rev. Lett.} \textbf{\bibinfo{volume}{50}},
  \bibinfo{pages}{1153} (\bibinfo{year}{1983}{\natexlab{b}}).

\bibitem[{\citenamefont{Affleck et~al.}(1987)\citenamefont{Affleck, Kennedy,
  Lieb, and Tasaki}}]{affleck1987}
\bibinfo{author}{\bibfnamefont{I.}~\bibnamefont{Affleck}},
  \bibinfo{author}{\bibfnamefont{T.}~\bibnamefont{Kennedy}},
  \bibinfo{author}{\bibfnamefont{E.~H.} \bibnamefont{Lieb}}, \bibnamefont{and}
  \bibinfo{author}{\bibfnamefont{H.}~\bibnamefont{Tasaki}},
  \bibinfo{journal}{Phys. Rev. lett.} \textbf{\bibinfo{volume}{59}},
  \bibinfo{pages}{799} (\bibinfo{year}{1987}).

\bibitem[{\citenamefont{Kennedy}(1990)}]{kennedy1990}
\bibinfo{author}{\bibfnamefont{T.}~\bibnamefont{Kennedy}},
  \bibinfo{journal}{J.~Phys.~Condens.~Matter} \textbf{\bibinfo{volume}{2}},
  \bibinfo{pages}{5737} (\bibinfo{year}{1990}).

\bibitem[{\citenamefont{Hagiwara et~al.}(1990)\citenamefont{Hagiwara,
  Katsumata, Affleck, Halperin, and Renard}}]{hagiwara1990}
\bibinfo{author}{\bibfnamefont{M.}~\bibnamefont{Hagiwara}},
  \bibinfo{author}{\bibfnamefont{K.}~\bibnamefont{Katsumata}},
  \bibinfo{author}{\bibfnamefont{I.}~\bibnamefont{Affleck}},
  \bibinfo{author}{\bibfnamefont{B.~I.} \bibnamefont{Halperin}},
  \bibnamefont{and} \bibinfo{author}{\bibfnamefont{J.~P.}
  \bibnamefont{Renard}}, \bibinfo{journal}{Phys. Rev. Lett.}
  \textbf{\bibinfo{volume}{65}}, \bibinfo{pages}{3181} (\bibinfo{year}{1990}).

\bibitem[{\citenamefont{Ng}(1994)}]{ng1994}
\bibinfo{author}{\bibfnamefont{T.-K.} \bibnamefont{Ng}},
  \bibinfo{journal}{Phys. Rev. B} \textbf{\bibinfo{volume}{50}},
  \bibinfo{pages}{555} (\bibinfo{year}{1994}).

\bibitem[{\citenamefont{Bi et~al.}(2013)\citenamefont{Bi, Rasmussen, and
  Xu}}]{xulinedefect}
\bibinfo{author}{\bibfnamefont{Z.}~\bibnamefont{Bi}},
  \bibinfo{author}{\bibfnamefont{A.}~\bibnamefont{Rasmussen}},
  \bibnamefont{and} \bibinfo{author}{\bibfnamefont{C.}~\bibnamefont{Xu}},
  \bibinfo{journal}{arXiv:1304.7272}  (\bibinfo{year}{2013}).

\bibitem[{\citenamefont{Xu}(2013{\natexlab{b}})}]{xuset}
\bibinfo{author}{\bibfnamefont{C.}~\bibnamefont{Xu}},
  \bibinfo{journal}{arXiv:1307.8131}  (\bibinfo{year}{2013}{\natexlab{b}}).

\bibitem[{\citenamefont{Kapustin}(2014{\natexlab{a}})}]{kapustin1}
\bibinfo{author}{\bibfnamefont{A.}~\bibnamefont{Kapustin}},
  \bibinfo{journal}{arXiv:1404.6659}  (\bibinfo{year}{2014}{\natexlab{a}}).

\bibitem[{\citenamefont{Kapustin}(2014{\natexlab{b}})}]{kapustin4}
\bibinfo{author}{\bibfnamefont{A.}~\bibnamefont{Kapustin}},
  \bibinfo{journal}{arXiv:1403.1467}  (\bibinfo{year}{2014}{\natexlab{b}}).

\bibitem[{\citenamefont{Kong and Wen}(2014)}]{kongwen}
\bibinfo{author}{\bibfnamefont{L.}~\bibnamefont{Kong}} \bibnamefont{and}
  \bibinfo{author}{\bibfnamefont{X.-G.} \bibnamefont{Wen}},
  \bibinfo{journal}{arXiv:1405.5858}  (\bibinfo{year}{2014}).

\bibitem[{\citenamefont{You and Xu}(2014{\natexlab{a}})}]{xu6d}
\bibinfo{author}{\bibfnamefont{Y.-Z.} \bibnamefont{You}} \bibnamefont{and}
  \bibinfo{author}{\bibfnamefont{C.}~\bibnamefont{Xu}},
  \bibinfo{journal}{arXiv:1410.6486}  (\bibinfo{year}{2014}{\natexlab{a}}).

\bibitem[{\citenamefont{Bi and Xu}(2015)}]{xubeyond}
\bibinfo{author}{\bibfnamefont{Z.}~\bibnamefont{Bi}} \bibnamefont{and}
  \bibinfo{author}{\bibfnamefont{C.}~\bibnamefont{Xu}},
  \bibinfo{journal}{arXiv:1501.02271}  (\bibinfo{year}{2015}).

\bibitem[{\citenamefont{Witten}(1984)}]{witten1984}
\bibinfo{author}{\bibfnamefont{E.}~\bibnamefont{Witten}},
  \bibinfo{journal}{Commun. Math. Phys.} \textbf{\bibinfo{volume}{92}},
  \bibinfo{pages}{455} (\bibinfo{year}{1984}).

\bibitem[{\citenamefont{Knizhnik and
  Zamolodchikov}(1984)}]{KnizhnikZamolodchikov1984}
\bibinfo{author}{\bibfnamefont{V.~G.} \bibnamefont{Knizhnik}} \bibnamefont{and}
  \bibinfo{author}{\bibfnamefont{A.~B.} \bibnamefont{Zamolodchikov}},
  \bibinfo{journal}{Nucl. Phys. B} \textbf{\bibinfo{volume}{247}},
  \bibinfo{pages}{83} (\bibinfo{year}{1984}).

\bibitem[{\citenamefont{Levine et~al.}(1983)\citenamefont{Levine, Libby, and
  Pruisken}}]{pruisken1}
\bibinfo{author}{\bibfnamefont{H.}~\bibnamefont{Levine}},
  \bibinfo{author}{\bibfnamefont{S.~B.} \bibnamefont{Libby}}, \bibnamefont{and}
  \bibinfo{author}{\bibfnamefont{A.~M.~M.} \bibnamefont{Pruisken}},
  \bibinfo{journal}{Phys. Rev. Lett.} \textbf{\bibinfo{volume}{51}},
  \bibinfo{pages}{1915} (\bibinfo{year}{1983}).

\bibitem[{\citenamefont{Levine et~al.}(1984)\citenamefont{Levine, Libby, and
  Pruisken}}]{pruisken2}
\bibinfo{author}{\bibfnamefont{H.}~\bibnamefont{Levine}},
  \bibinfo{author}{\bibfnamefont{S.~B.} \bibnamefont{Libby}}, \bibnamefont{and}
  \bibinfo{author}{\bibfnamefont{A.~M.~M.} \bibnamefont{Pruisken}},
  \bibinfo{journal}{Nucl. Phys. B} \textbf{\bibinfo{volume}{240}},
  \bibinfo{pages}{30, 49, 71} (\bibinfo{year}{1984}).

\bibitem[{\citenamefont{Xu and Ludwig}(2013)}]{xuludwig2013}
\bibinfo{author}{\bibfnamefont{C.}~\bibnamefont{Xu}} \bibnamefont{and}
  \bibinfo{author}{\bibfnamefont{A.~W.~W.} \bibnamefont{Ludwig}},
  \bibinfo{journal}{Phys. Rev. Lett.} \textbf{\bibinfo{volume}{110}},
  \bibinfo{pages}{200405} (\bibinfo{year}{2013}).

\bibitem[{\citenamefont{Qi et~al.}(2008)\citenamefont{Qi, Hughes, and
  Zhang}}]{qi2008}
\bibinfo{author}{\bibfnamefont{X.-L.} \bibnamefont{Qi}},
  \bibinfo{author}{\bibfnamefont{T.~L.} \bibnamefont{Hughes}},
  \bibnamefont{and} \bibinfo{author}{\bibfnamefont{S.-C.} \bibnamefont{Zhang}},
  \bibinfo{journal}{Phys. Rev. B} \textbf{\bibinfo{volume}{78}},
  \bibinfo{pages}{195424} (\bibinfo{year}{2008}).

\bibitem[{\citenamefont{Essin et~al.}(2009)\citenamefont{Essin, Moore, and
  Vanderbilt}}]{mooretheta}
\bibinfo{author}{\bibfnamefont{A.~M.} \bibnamefont{Essin}},
  \bibinfo{author}{\bibfnamefont{J.~E.} \bibnamefont{Moore}}, \bibnamefont{and}
  \bibinfo{author}{\bibfnamefont{D.}~\bibnamefont{Vanderbilt}},
  \bibinfo{journal}{Phys. Rev. Lett.} \textbf{\bibinfo{volume}{102}},
  \bibinfo{pages}{146805} (\bibinfo{year}{2009}).

\bibitem[{\citenamefont{Lieb et~al.}(1961)\citenamefont{Lieb, Schultz, and
  Mattis}}]{LSM}
\bibinfo{author}{\bibfnamefont{E.~H.} \bibnamefont{Lieb}},
  \bibinfo{author}{\bibfnamefont{T.~D.} \bibnamefont{Schultz}},
  \bibnamefont{and} \bibinfo{author}{\bibfnamefont{D.~C.}
  \bibnamefont{Mattis}}, \bibinfo{journal}{Ann. Phys.}
  \textbf{\bibinfo{volume}{16}}, \bibinfo{pages}{407} (\bibinfo{year}{1961}).

\bibitem[{\citenamefont{Kitaev}(2003)}]{kitaev2003}
\bibinfo{author}{\bibfnamefont{A.~Y.} \bibnamefont{Kitaev}},
  \bibinfo{journal}{Ann. Phys.} \textbf{\bibinfo{volume}{303}},
  \bibinfo{pages}{1} (\bibinfo{year}{2003}).

\bibitem[{\citenamefont{Grover and Senthil}(2008)}]{groversenthil}
\bibinfo{author}{\bibfnamefont{T.}~\bibnamefont{Grover}} \bibnamefont{and}
  \bibinfo{author}{\bibfnamefont{T.}~\bibnamefont{Senthil}},
  \bibinfo{journal}{Phys. Rev. Lett.} \textbf{\bibinfo{volume}{100}},
  \bibinfo{pages}{156804} (\bibinfo{year}{2008}).

\bibitem[{\citenamefont{Chen et~al.}(2014{\natexlab{b}})\citenamefont{Chen,
  Burnell, Vishwanath, and Fidkowski}}]{lukaszsemion}
\bibinfo{author}{\bibfnamefont{X.}~\bibnamefont{Chen}},
  \bibinfo{author}{\bibfnamefont{F.~J.} \bibnamefont{Burnell}},
  \bibinfo{author}{\bibfnamefont{A.}~\bibnamefont{Vishwanath}},
  \bibnamefont{and}
  \bibinfo{author}{\bibfnamefont{L.}~\bibnamefont{Fidkowski}},
  \bibinfo{journal}{arXiv:1403.6491}  (\bibinfo{year}{2014}{\natexlab{b}}).

\bibitem[{\citenamefont{Liu et~al.}(2011)\citenamefont{Liu, Chen, and
  Wen}}]{1dd2h}
\bibinfo{author}{\bibfnamefont{Z.-X.} \bibnamefont{Liu}},
  \bibinfo{author}{\bibfnamefont{X.}~\bibnamefont{Chen}}, \bibnamefont{and}
  \bibinfo{author}{\bibfnamefont{X.-G.} \bibnamefont{Wen}},
  \bibinfo{journal}{Phys. Rev. B} \textbf{\bibinfo{volume}{84}},
  \bibinfo{pages}{195145} (\bibinfo{year}{2011}).

\bibitem[{\citenamefont{Burnell et~al.}(2013)\citenamefont{Burnell, Chen,
  Fidkowski, and Vishwanath}}]{fionachen}
\bibinfo{author}{\bibfnamefont{F.~J.} \bibnamefont{Burnell}},
  \bibinfo{author}{\bibfnamefont{X.}~\bibnamefont{Chen}},
  \bibinfo{author}{\bibfnamefont{L.}~\bibnamefont{Fidkowski}},
  \bibnamefont{and}
  \bibinfo{author}{\bibfnamefont{A.}~\bibnamefont{Vishwanath}},
  \bibinfo{journal}{arXiv:1302.7072}  (\bibinfo{year}{2013}).

\bibitem[{\citenamefont{Xu and Sachdev}(2009)}]{xusachdev}
\bibinfo{author}{\bibfnamefont{C.}~\bibnamefont{Xu}} \bibnamefont{and}
  \bibinfo{author}{\bibfnamefont{S.}~\bibnamefont{Sachdev}},
  \bibinfo{journal}{Phys. Rev. B} \textbf{\bibinfo{volume}{79}},
  \bibinfo{pages}{064405} (\bibinfo{year}{2009}).

\bibitem[{\citenamefont{You and Xu}(2014{\natexlab{b}})}]{youxuinversion}
\bibinfo{author}{\bibfnamefont{Y.-Z.} \bibnamefont{You}} \bibnamefont{and}
  \bibinfo{author}{\bibfnamefont{C.}~\bibnamefont{Xu}}, \bibinfo{journal}{Phys.
  Rev. B} \textbf{\bibinfo{volume}{90}}, \bibinfo{pages}{245120}
  (\bibinfo{year}{2014}{\natexlab{b}}).

\bibitem[{\citenamefont{Wang and Levin}(2014)}]{levinloop}
\bibinfo{author}{\bibfnamefont{C.}~\bibnamefont{Wang}} \bibnamefont{and}
  \bibinfo{author}{\bibfnamefont{M.}~\bibnamefont{Levin}},
  \bibinfo{journal}{Phys. Rev. Lett.} \textbf{\bibinfo{volume}{113}},
  \bibinfo{pages}{080403} (\bibinfo{year}{2014}).

\bibitem[{\citenamefont{Bi et~al.}(2014)\citenamefont{Bi, You, and
  Xu}}]{xuloop}
\bibinfo{author}{\bibfnamefont{Z.}~\bibnamefont{Bi}},
  \bibinfo{author}{\bibfnamefont{Y.-Z.} \bibnamefont{You}}, \bibnamefont{and}
  \bibinfo{author}{\bibfnamefont{C.}~\bibnamefont{Xu}}, \bibinfo{journal}{Phys.
  Rev. B} \textbf{\bibinfo{volume}{90}}, \bibinfo{pages}{081110}
  (\bibinfo{year}{2014}).

\end{thebibliography}
\end{document}